\documentclass[printer]{aa}  

\usepackage{graphicx}
\usepackage{amsmath}
\usepackage{amssymb}
\usepackage{txfonts}
\usepackage{hyperref}
\usepackage{cprotect} 
\usepackage{multirow}

\def\lum{erg\,s$^{-1}$}
\def\fint{erg\,cm$^{-2}$\,s$^{-1}$}
\def\flux{erg\,cm$^{-2}$\,s$^{-1}$}

\def\xmmn{{\it XMM-Newton}}
\def\sfr{M$_\odot$yr$^{-1}$}

\def\psfr{/M$_\odot$yr$^{-1}$}

\begin{document}

   \title{An expanded ultraluminous X-ray source catalogue\thanks{Based on
  observations obtained with \xmmn, an ESA science mission with instruments and contributions directly funded by ESA Member States and NASA.}}

   \author{M. C. i Bernadich\inst{1,2,3}
          \and
          A. D. Schwope\inst{1}
          \and
          K. Kovlakas\inst{4,5}
          \and
          A. Zezas\inst{4,5,6}
          \and
          I. Traulsen\inst{1}
          }

   \institute{Leibniz-Institut f\"ur Astrophysik Potsdam (AIP), An der Sternwarte 16, 14482 Potsdam, Germany
   \and
   Observational Astrophysics, Division of Astronomy and Space Physics, Department of Physics and Astronomy, Uppsala University, Box 516, 75120 Uppsala, Sweden
   \and
   Max Plank Institut für Radioastronomie (MPIfR), Auf dem Hügel 69, 53121 Bonn, Germany
   \and
   Physics Department, University of Crete, GR 71003, Heraklion, Greece
   \and
   Institute of Astrophysics, Foundation for Research and Technology – Hellas, GR-70013 Heraklion, Greece
   \and
   Harvard-Smithsonian Center for Astrophysics, 60 Garden Street, Cambridge, MA 02138, USA
             }

   \date{Received ; accepted }

\abstract{Ultraluminous X-ray sources (ULXs) are non-nuclear, extragalactic, point-like X-ray sources whose luminosity exceeds that of the Eddington limit of an accreting stellar-mass black hole ($M_\textrm{BH}<10\textrm{M}_\odot$, $L_\textrm{X}>10^{39}$\,\lum). They are excellent laboratories for extreme accretion physics, probes for the star formation history of galaxies, and constitute precious targets for the search of intermediate-mass black holes. As the sample size of X-ray data from modern observatories such as \textit{XMM-Newton} and \textit{Chandra} increases, producing extensive catalogues of ULXs and studying their collective properties has become both a possibility and a priority.}
{We build a clean, updated ULX catalogue based on one of the most recent \textit{XMM-Newton} X-ray serendipitous survey data releases, \textit{4XMM-DR9}, and the most recent and exhaustive catalogue of nearby galaxies, \textit{HECATE}. We perform a preliminary population study to test if the properties of the expanded \textit{XMM-Newton} ULX population are consistent with previous findings.}
{We perform positional cross-matches between \textit{XMM-Newton} sources and \textit{HECATE} objects to identify host galaxies. We filter out known foreground and background sources and other interlopers by finding counterparts in external catalogues and databases such as \textit{Gaia DR2}, \textit{SSDS}, \textit{PanSTARRS1}, \textit{NASA/IPAC Extragalactic Database} and \textit{SIMBAD}. Manual inspection of image data from \textit{PanSTARRS1} and the \textit{NASA/IPAC} is occasionally performed to investigate the nature of some individual sources. We use distance and luminosity arguments to identify ULX candidates. Source parameters from \textit{4XMM-DR9}, galaxy parameters from \textit{HECATE}, and variability parameters from \textit{4XMM-DR9s} are used to study the spectral, abundance and variability properties of the updated \textit{XMM-Newton} ULX sample.}
{We identify 779 ULX candidates, 94 of which hold $L_\textrm{X}\gtrsim5\times10^{40}$\,\lum. We find that spiral galaxies are more likely to host ULXs. For early-spiral galaxies we also find that the number of ULX candidates per star forming rate that is consistent with previous studies, while we also attest the existence of a significant ULX population in elliptical and lenticular galaxies. Candidates hosted by late-type galaxies tend to present harder spectra and to undergo more and more extreme inter-observation variability than the ones hosted by early-type galaxies. $\sim$30 candidates with $L_\textrm{X}>10^{41}$\,\lum\ are also identified, constituting the most interesting candidates for intermediate-mass black hole searches.}
{We have built the largest ULX catalogue to this date. Our results regarding the spectral and abundance properties of ULXs confirm the findings made by previous studies based on \textit{XMM-Newton} and \textit{Chandra} data, while our population-scale study on variability properties is unprecedented. Our study, however, provides limited insight on the properties of the bright ULX candidates ($L_\textrm{X}\gtrsim5\times10^{40}$\,\lum) due to the small sample size. The expected growth of X-ray catalogues and potential future follow-ups will aid in drawing a more clear picture.}
   \authorrunning{Bernadich et al.}   
   \maketitle    

%

\section{Introduction}

   Ultraluminous X-ray sources (ULXs) are extragalactic, point-like, X-ray sources whose luminosity exceeds that of the Eddington limit of an accreting stellar-mass black hole (StBH, $M_\textrm{BH}<10\textrm{M}_\odot$, $L_\textrm{Edd}{\approx}10^{39}$\,\lum), and that are not the central source of their host galaxy. These objects have been the subject of active research since the advent of X-ray astronomy (see \citealt{Kaaret2017} for a review) and, for a time, due to the scaling of the Eddington limit with mass, it was believed that they hosted accreting BHs of masses between those of StBHs and supermassive BHs. Nowadays, they are more commonly associated with super-Eddington accretion onto common compact objects such as StBH and neutron stars (NSs), with a few exceptions where BHs of higher mass need to be invoked \citep{Bachetti2016}.

   The current understanding comes from three main lines of evidence. Firstly, predictions of the formation rate for intermediate-mass BHs (IMBHs, $100\textrm{M}_\odot<M_\textrm{BH}<10^{6}\textrm{M}_\odot$) with donor companions fall short of explaining the observed abundances of ULXs \citep{King2004,Madhusudhan2006}. Secondly, ULXs are most commonly associated with star-forming galaxies \citep{Swartz2011}, or even young elliptical galaxies with recent star formation events, rather than old elliptical ones with no star formation \citep{KimFabbiano2004,KimFabbiano2010}. This is consistent with ULXs belonging to a high-luminosity extrapolation of the X-ray binary population, whose abundances correlate with the star-formation rate (SFR) and the stellar mass ($M_*$) of their host galaxies \citep{Gilfanov2004,Grimm2003,Mineo2012}. And finally, but perhaps most importantly, the direct observation of regular pulsations in some ULXs unmistakably points towards accretion onto a NS in binary systems. Famous examples of neutron star-powered ULXs (NS-ULXs) are M82 X-2 \citep{Bachetti2014}, NGC 7793 P13 \citep{Furst2016,Israel2016}, NGC 5907 ULX1 \citep{Israel2017} or NGC 300 ULX1 \citep{Carpano2018}. Since some of these objects present $L_\textrm{X}>10^{40}$\,\lum, they are clear cases of super-Eddington accretion. In hand with these observations, population synthesis models even suggest that NS-ULXs constitute a significant fraction of the ULX population \citep{Wiktorowicz2017}.
   
    The prospect of super-Eddington in ULXs accretion has sparked a lot of interest for the modelling of ULX systems. Many works have shown that sufficiently extreme accretion rates to power ULXs can be achieved during certain phases of regular HMXB and LMXB evolution. In fact, these rates can explain most of the ULX population \citep{King2002,Rappaport2005,Wiktorowicz2015,Pavlovskii2017}. Yet, many aspects of the physics of super-Eddington accretion itself are poorly understood, specially in cases where the material falls on top of strongly magnetized neutron stars, and very rich theoretical research is being performed to model this phenomenon (see \citealt{Kaaret2017} for a comprehensive discussion of this topic, which otherwise lies outside of the scope of this introduction).
    
    Nonetheless, ULXs are still good sources where to look for IMBHs, as a handful of specially bright candidates still remain. ESO 243-49 HLX-1 has been detected with both luminosities of $L_\textrm{X}{\approx}10^{42}$~\lum and spectral state transitions typical of sub-Eddington Galactic BH binaries \citep[GBHBs,][]{RemillardMcClintock2006}, being consistent with a mass of $\sim$$10^{3}$~M$_\odot$ \citep{Servillat2011}. M82 X-1 showcases a similar behavior \citep{Kong2007}, with the addition of quasi-periodic oscillations also characteristic from GBHBs (QPO, Remillard \& McClintock \citeyear{RemillardMcClintock2006}) and that grant it a mass estimate of $400~\textrm{M}_\odot$ \citep{Pasham2014}. And just like in GBHBs, powerful radio jets are also present in the IMBH candidates ESO 243-49 HLX-1 \citep{Webb2012,Cseh2015} and NGC 2276-3c \citep{Mezcua2015}.
   
   The relevance of ULXs as proxies of recent star formation, in accretion physics and in IMBHs searches has been so far exposed. However, ULX studies are limited by their small numbers and large distances. For this reason, a lot of effort has been put into building compilations of ULX candidates from X-ray surveys. Early works relied on \textit{ROSAT} data \citep[e.g.][]{RobertsWarwick2000,LiuGregman2005,Liu2006,ColbertPtak2002}, yielding up to $\sim$100 candidates. More recent ones have been based on \textit{XMM-Newton} and \textit{Chandra} data. \textit{XMM-Newton}-based catalogues have been highly successful, yielding up to $\sim$ 300 ULX candidates \citep{Walton2011,Earnshaw2019}, but they are somewhat hampered by the low angular resolution of the observatory. On the other hand, works based on the \textit{Chandra} telescope have provided better identification of individual sources thanks to its crisp angular resolution \citep[e.g.][]{Swartz2004,Swartz2011,Wang2016,Kovlakas2020}.
   
   Of particular interest to us are the works of \cite{Earnshaw2019} and \cite{Kovlakas2020}, since they both constructed ULXs samples from the largest and most recent \textit{XMM-Newton} and \textit{Chandra} samples available at the time. \cite{Earnshaw2019} is, in fact, our main predecessor and inspiration regarding methodology. They identify 384 ULX candidates within the fourth \textit{XMM-Newton} data release\footnote{\url{https://xmmssc-www.star.le.ac.uk/Catalogue/3XMM-DR4/UserGuide_xmmcat.html}} \citep[\textit{3XMM-DR4},][]{Rosen2016}, using the \textit{Third Reference Catalogue of Bright Galaxies} \citep[\textit{RC3},][]{deVaucouleurs1991} and the \textit{Catalogue of Neighbouring Galaxies} \citep{Karachentsev2004} as references for the host galaxies. Their study focuses mostly on the spectral properties of ULXs, and they find that ULXs tend to be somewhat harder in late-type galaxies, and that their hardness ratios in general alike to those of X-ray binaries below the Eddington threshold. \cite{Kovlakas2020} find 629 ULX candidates within the \textit{Chandra Source Catalog 2.0}\footnote{\url{https://cxc.harvard.edu/csc2/}} (\textit{CSC2}) using the \textit{Heraklion Extragalactic CATaloguE} \citep[\textit{HECATE}][]{Kovlakas2020b} as the reference for host galaxies, which comes with valuable information such as the SFR and $M_*$. Their study focuses heavily on the scaling properties of the ULX content of galaxies with the SFR and $M_*$ parameters, and finds that SFR is the determinant parameter for late-type galaxies while $M_*$ rules over the abundance in late-type galaxies, in accordance with recent population synthesis models \citep{Wiktorowicz2017}.

Now, the ninth \textit{XMM-Newton} data release\footnote{\url{http://xmmssc.irap.omp.eu}} \citep[\textit{4XMM-DR9},][]{Webb2019}, its stacked version \citep[\textit{4XMM-DR9s},][]{Traulsen2020} and the \textit{HECATE} catalogue are available to us, making it possible to build the largest catalogue of extragalactic, non-nuclear, point-like X-ray sources based on \textit{XMM-Newton} data. We then use the catalogue to preliminarily explore the distribution of ULX spectral and variability properties, and the dependence of ULX abundances galaxies with the SFR and $M_*$ on \textit{XMM-Newton}'s sky.

This paper is organized as follows: in the following section we explain the specifics of selection in archival data, in Section~\ref{pipeline} we describe our methodology for identifying ULX candidates and other interloping objects, in Section~\ref{resultsSection} we expose our main results and in Section~\ref{discussion} we discuss some of the limitations and implications of our work and compare it with previous ones.

\section{Data samples}\label{dataSamples}

\subsection{The X-ray samples}\label{XraySample}

We used the Fourth \textit{XMM-Newton} Serendipitous Source Catalog, Ninth Data Release, \citep[\textit{4XMM-DR9},][]{Webb2019} as the basis for our ULX catalogue. It lists 810\,795 individual detections of 550\,124 unique sources discovered across 11\,204 observations, and it represents an increase of 177\,396 sources with respect to \textit{3XMM-DR4}, the resource used in the previous ULX study based in \textit{XMM-Newton} data \citep{Earnshaw2019}.
The columns we used during the construction are:
\begin{itemize}
   \item \verb|DETID| and \verb|SRCID|, the identification number of detections and sources,
   \item \verb|SC_RA|, \verb|SC_DEC| and \verb|SC_POSERR|, the source J2000.0 sky coordinates and their positional uncertainty,
   \item \verb|SC_EXTENT|, the measured extension of the source,
   \item \verb|SC_DET_ML|, the detection likelihood of sources, taken from the highest value of \verb|DET_ML| among their detections,
   \item \verb|EP_8_FLUX| and \verb|EP_8_FLUX_ERR|, the measured X-ray flux of detections and its uncertainty in the 0.2$-$12\,keV band, derived from the EPIC photon counts,
   \item \verb|SC_EP_8_FLUX| and \verb|SC_EP_8_FLUX_ERR|, the averaged X-ray flux of sources and its uncertainty in the 0.2$-$12\,keV band, derived from the EPIC photon counts,
   \item \verb|SC_HR|$i$, where $i$ runs from 1 to 4, the source hardness ratios from the source count rates in the respective energy bands,
   \item and \verb|SC_SUM_FLAG|, the summary quality flag of a unique source, which runs from 0 to 5. 0 means that all detections of the source are trustworthy, while 5 means that at least one detection is most likely spurious. Its value is given by the highest value of \verb|SUM_FLAG| among all detections of the source.
\end{itemize}

Since the \textit{XMM-Newton} FWHM resolution is of 6\arcsec\ \citep{ESA2019}, we included only sources with $\verb|SC_EXTENT|<6\arcsec$ to work with point-like sources only. We also considered only sources with $\verb|SC_DET_ML|>8$. As these parameters are based on the worst result for all the detections of a source, it can happen that an otherwise point-like source is counted as extended if in one of its detections it is measured as such. We nevertheless excluded them from the catalogue because other source parameters that consist in averaged detection parameters are most likely unreliable. This left 452\,602 sources available for the catalogue.

Additionally, we used the stacked version of \textit{4XMM-DR9}. This is the second \textit{XMM-Newton} serendipitous source catalogue built from overlapping observations \citep[\textit{4XMM-DR9s},][]{Traulsen2020}, which compiles variability information for 218\,283 sources detected in 6\,604 overlapping \textit{XMM-Newton} observations, an improvement of 146\,332 sources with respect to the first version \citep{Traulsen2019}. The strength of \textit{4XMM-DR9s} is that, from the overlapping observations, extra source variability parameters that are not included in the bare \textit{4XMM-DR9} are computed. From it, we used the following columns:
\begin{itemize}
   \item \verb|SRCID|, the source identification number,
   \item \verb|RA|, \verb|DEC| and \verb|RADEC_ERR|, the source J2000.0 sky coordinates and their uncertainty,
   \item \verb|N_CONTRIB|, the number of times a source contributes to the computation of variability, usually equal to the number of times it has been observed,
   \item \verb|VAR_PROB|, the probability of a source \textit{not} showing inter-observation variability, computed from the reduced $\chi^2$ of EPIC flux variability,
   \item \verb|FRATIO|, the $F_\textrm{max}/F_\textrm{min}$ ratio between the highest and lowest observed fluxes in a single source.
\end{itemize}

\subsection{The galaxy sample}\label{GraySample}

We used the Heraklion Extragalactic CATaloguE \citep[\textit{HECATE},][submitted]{Kovlakas2020b} compilation of 204\,733 galaxies within 200\,Mpc. Built from the HyperLEDA catalogue \citep{Makarov2014} and other databases such as \textit{NED}, SDSS and 2MASS, it is a much more complete galaxy compilation than the \textit{Third Reference Catalogue of Galaxies} \citep[\textit{RC3},][]{deVaucouleurs1991}, traditionally used during previous ULX studies \citep{Swartz2011,Wang2016,Earnshaw2019}. A more detailed description of its contents is available in Kovlakas et al. \citeyear{Kovlakas2020b}, but here we mention the columns we used:
\begin{itemize}
   \item \verb|PGC|, the Principal Galaxy Catalogue identification number, originating from the \textit{Principal Catalogue of Galaxies} \citep{Paturel1985} and still used in HyperLEDA,
   \item \verb|RA| and \verb|DEC|, the J2000.0 coordinates of the galactic center,
   \item \verb|R1| and \verb|R2|, the minor and major D25 isophotal radii,
   \item \verb|PA|, the North-to-East position angle of the major axis,
   \item \verb|D| and \verb|D_ERR|, the galaxy distance and its uncertainty,
   \item \verb|T|, the Hubble Type value ($T_\textrm{H}$) of a galaxy,
   \item \verb|logSFR_HEC|, the decimal logarithm of the SFR estimate,
   \item and \verb|logM_HEC|, the decimal logarithm of $M_*$.
\end{itemize}

The SFR values in the \textit{HECATE} are based on infrared calibrations, which tend to overestimate the SFR in early-type galaxies \citep{Kovlakas2020b}. Therefore, SFR values were only inspected for galaxies with $T_\textrm{H}\geq0$.

\subsection{The interloper samples}

Our X-ray catalogue is inevitably populated by a fraction of non-ULX contaminating objects such as background AGNs or foreground stars. Therefore, we built a filtering pipeline (see section \ref{filter}) to identify interlopers in external catalogues and databases of already known objects. These consist of: the second \textit{Gaia} data release \citep[\textit{GaiaDR2},][]{Gaia2018}, the \textit{Tycho-2} catalogue of bright stars \citep[\textit{Tycho2},][]{Hog2000}, the 14th \textit{Sloan Digital Sky Survey} data release \citep[\textit{SDSS-DR14},][]{Blanton2017}, the 13th edition of the Véron-Cetty \& Véron catalogue of QSOs and AGNs \citep[\textit{VéronQSO},][]{Veron2010}, the \textit{SIMBAD} database \citep{Wegner2000}, the \textit{Panoramic Survey Telescope and Rapid Response System} database\footnote{https://panstarrs.stsci.edu} \citep[\textit{PanSTARRS1},][]{Flewelling2016} from the \textit{PanSTARRS1} surveys \citep{Chambers2016} and the NASA/IPAC Extragalactic Database\footnote{\url{https://ned.ipac.caltech.edu/}} (\textit{NED}).

All the cross-matches were performed using the TOPCAT\footnote{\url{http://www.star.bris.ac.uk/mbt/topcat/}} and STILTS\footnote{\url{http://www.star.bris.ac.uk/~mbt/stilts/}} software tools to manipulate the tables in all of the following steps \citep{Taylor2005}.

\section{Methods}\label{pipeline}

\subsection{The automatic filtering process}\label{filter}

\textbf{Correlation of sources with galaxies.} We selected all \textit{4XMM-DR9} sources that overlapped with the isophotal ellipses of \textit{HECATE} within their positional uncertainty. Catalogue entries matched to galaxies with available measurements of \verb|R1|, \verb|R2| and \verb|PA| were labeled with $\verb|MATCH_FLAG|=0$. When \verb|PA| was unknown, detections of sources matched with the minor isophotal circle were labelled as $\verb|MATCH_FLAG|=1$, while detections of sources within the annulus drawn by \verb|R1| and \verb|R2| were given $\verb|MATCH_FLAG|=2$. When both \verb|PA| and \verb|R2| were unlisted, sources matched to the circle drawn by \verb|R1| were labeled as $\verb|MATCH_FLAG|=3$. Additionally, we stored in \verb|n_Galaxies| the number of galaxies a source has been matched with. Detections of sources with $\verb|n_Galaxies|>1$ are listed more than once and treated independently for each galaxy, being identifiable by the unique \verb|DET_PGC_ID| and \verb|SRC_PGC_ID| detection/source-galaxy identity numbers. This way, a comprehensive catalogue of 50\,446 entries was built, out of which 49\,816 have $\verb|n_Galaxies|=1$ and 48\,206 present $\verb|MATCH_FLAG|=0$.

\begin{figure}[t]
\centering
\includegraphics[width=0.9\columnwidth]{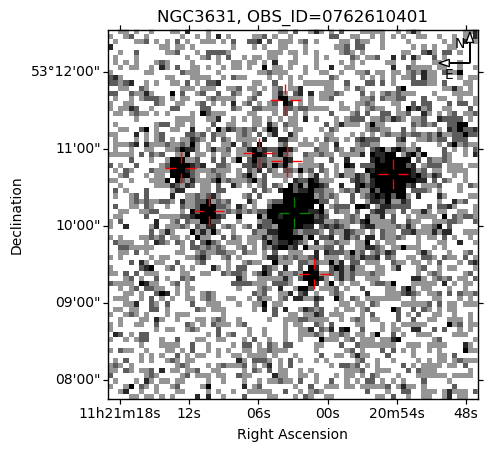}
\end{figure}

\begin{figure}[t]
\centering
\includegraphics[width=0.9\columnwidth]{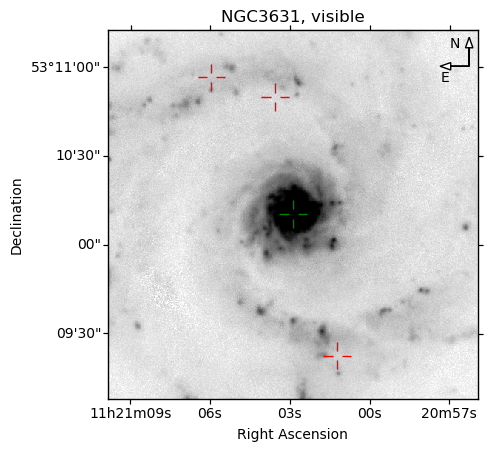}
\caption{\textbf{Top:} \textit{XMM-Newton} observation 0762610401, including galaxy \object{NGC 3631} with all detected sources. \textbf{Bottom:} optical Pan-STARRS1 \citep{Chambers2016} image of the same galaxy. Non-nuclear sources are highlighted with red crosses, while a green cross signals the central source. The image in the visible range covers a slightly smaller sky area including the four most central sources from the X-ray image.}
\label{NGC3631}
\end{figure}

\textbf{Identification of central sources.} At low redshifts, AGNs typically present luminosities of $L_\textrm{X}>10^{41}$\,\lum\ \citep{Brandt2015}. However, some have been observed to overlap with the ULX luminosity regime \citep{Ho2001,Eracleous2002,Ghosh2008,Zhang2009}. Therefore, we identified all sources whose position overlapped within three times of their uncertainty with a circle of radius 3" around the center of their host galaxy. 3\,658 entries in our catalogue were thus labeled as central sources. We also checked for sources with more than one potential host galaxy ($\verb|n_Galaxies|>1$) that had been flagged as central-source candidates in one of their iterations, and flagged them as central-source candidates on all of their other galaxy associations to indicate their potentially nuclear nature. This highlighted 137 further entries, all of which are composed by optically coincident galaxies, whether related or not. Figure~\ref{NGC3631} illustrates how essential it is to identify these central sources before classifying them as ULXs, as otherwise they would easily slip through and contaminate our sample.

\textbf{Identification of foreground stellar objects.}

X-ray emission from stars is ubiquitous across the Hertzsprung-Russell diagram and has manifold reasons, be it active coronae, hot stellar winds or binary accretion, which can lead to the misclassification of Galactic objects as extragalactic ULXs. We performed a cross-match of unflagged objects with \textit{GaiaDR2} making use of the query tool provided by CDS, Strasbourg, and then with \textit{Tycho2}, to identify potential contaminating foreground stars. In both cases, a positional overlap within three times the positional uncertainty was required on both sides of the cross-match. Assuming that all matched objects were stars, a very stringent constraint of $\log(F_\textrm{X}/F_\textrm{V})<-2.2$ was used to decide whether the matched object can explain the X-ray luminosity of the \textit{XMM-Newton} source, where $F_\textrm{X}$ is the source X-ray flux provided by \verb|SC_EP_8_FLUX| and $F_\textrm{V}$ its flux in optical light. This limit is established in consistency with the work of \cite{Freund2018}, were $\log(F_\textrm{X}/F_\textrm{BOL})=-2.2$ is considered the maximal luminosity ratio for early-type stars, the ones presenting the largest share of X-ray luminosity. However, being the bolometric flux $F_\textrm{BOL}$ unavailable in most existing catalogues, we approximated it as the optical flux of the star, $F_\textrm{V}$. We used the formula from \cite{Maccacaro1988} to write \begin{equation}\label{stars}\log\left(\frac{F_\textrm{X}}{F_\textrm{V}}\right)=\log{F_\textrm{X}}+\frac{m_\textrm{V}}{2.5}+5.37<-2.2\textrm{,}\end{equation}
where $m_\textrm{V}$ is the optical magnitude of the studied object. In cases where $m_\textrm{V}$ is not listed, we used the G-band magnitude $m_\textrm{G}$ instead, which typically has the same value at the V-band magnitude for bright enough sources. From \textit{GaiaDR2}, we used the listed G-band magnitude values \verb|Gmag|, and for \textit{Tycho2} we used the VT-magnitude measurements \verb|VTmag|. This resulted in 2\,257 entries flagged as stars from \textit{GaiaDR2}; and 96 entries from \textit{Tycho2}.

ULXs optical counterparts are typically very faint in the optical, presenting $m_\textrm{V}\gtrsim21$ \citep{Kaaret2017}, which leads to $\log(F_\textrm{X}/F_\textrm{V})\gtrsim 0$ for a ULX candidate with $L_\textrm{X}=10^{39}$\,\lum\ at a distance of 20 Mpc. Therefore, we are safe from accidentally disregarding genuine ULX candidates as stellar objects. Objects with $0>\log(F_\textrm{X}/F_\textrm{V})>-2.2$ were considered in an upcoming section of the pipeline.

\textbf{Identification of background QSOs.} 
From the 2-10 keV $\log{N}-\log{S}$ distribution above a flux of $10^{-14}$~\flux\,presented in \cite{Mateos2008}, we expect a background source density of 300~deg$^{-2}$ in the field. With an accumulated galaxy sky area outside of the local group ($D>1$~Mpc) of 2.25~deg$^{2}$, this implies that around 670 background contaminants lie in the line of sight of galaxies where we expect to find most of our ULX candidates. This clearly motivates the need to identify such objects in currently available catalogues. Therefore, all remaining unflagged objects found to overlap within three times their positional uncertainties during the cross-match with the \textit{SDSS-DR14} and \textit{VéronQSO} were flagged as QSOs. The first query was performed with \textit{SDSS-DR14} and highlighted 140 entries, while the second one highlighted 135 more.

\textbf{Identification of miscellaneous objects in \textit{SIMBAD}.} We cross-matched our remaining unflagged objects with the \textit{SIMBAD} database with the query tool provided by CDS, Strasbourg. In this step of the pipeline, sources were treated differently depending on the nature of the matched objects, indicated as $\verb|main_type|$ in the \textit{SIMBAD} catalogue\footnote{\url{http://simbad.u-strasbg.fr/simbad/sim-display?data=otypes}}. Every object overlapping within three times their positional uncertainty and with either $\verb|main_type|=$\textit{``Star''} or with their \verb|main_type| containing the ``*'' symbol, and that whose optical magnitude \verb|Vmag| holds equation (\ref{stars}) was flagged as a stellar contaminant. Objects with $\verb|main_type|=$\textit{``AGN''}, \textit{``AGN$\_$Candidate''}, \textit{``QSO''}, \textit{``QSO$\_$Candidate''} or \textit{``SN''} were also flagged, regardless of their optical magnitude. Supernovae in particular were highlighted as their emission is dominated from the expanding envelope. This step highlighted 5\,745 entries of the catalogue.

\textbf{Identification of miscellaneous optical objects.} As previously stated, ULXs optical counterparts typically present $m_\textrm{V}\gtrsim21$ \citep{Kaaret2017}, which leads to $\log(F_\textrm{X}/F_\textrm{V})\gtrsim 0$ for a ULX candidate with $L_\textrm{X}=10^{39}$\,\lum\ at a distance of 20 Mpc. Objects with larger optical luminosity are usually AGNs or foreground stars. Sources with $\log(F_\textrm{X}/F_\textrm{VOL})<-2.2$ were already labelled as stars in a previous section of the pipeline. In this section, we looked for optically bright objects of extragalactic origin. We performed a cross-match with the query tool provided by CDS, Strasbourg, with the \textit{PanSTARRS1} catalogue to find all objects that overlapped within a radius of three times their positional uncertainties and that held
\begin{equation}\label{optical objects}\log\left(\frac{F_\textrm{X}}{F_\textrm{V}}\right)=\log{F_\textrm{X}}+\frac{m_\textrm{V}}{2.5}+5.37<0\textrm{,}\end{equation}
to highlight all sources brighter in the optical than in the X-rays, using the G-band magnitude \verb|gmag| as an estimate for $m_\textrm{V}$. Up to 3\,151 entries were flagged as \textit{PanSTARRS1} extragalactic objects.

\begin{figure}[t!]
\centering
\includegraphics[width=0.95\columnwidth]{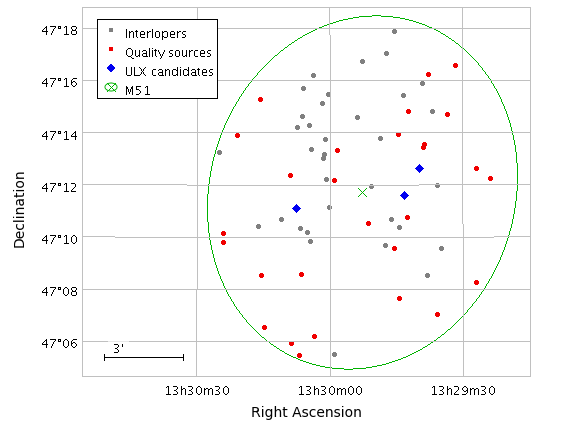}
\caption{TOPCAT astrometric map of \object{M51} showcasing all \textit{4XMM-DR9} sources within its isophotal ellipse. The interlopers are composed by \textit{PanSTARRS1} and \textit{SIMBAD} objects, and all of them but one stay below the ULX luminosity regime if assumed at the same distance as the others.}
\label{M51}
\end{figure}

\subsection{Identification of ULX candidates}\label{ULXcandidates}

To identify ULX candidates, we computed individual detection and average source luminosities from the X-ray fluxes as listed in \textit{4XMM-DR9} (Section~\ref{XraySample}) and from the distances as listed in \textit{HECATE} (Section~\ref{GraySample}), along with the corresponding error propagation. These are listed as \verb|Luminosity| and \verb|LuminosistyErr| for detections and \verb|SC_Luminosity| and \verb|SC_LuminosistyErr| for sources. In our catalogue, we consider any X-ray source to be within the ULX luminosity regime if:
\begin{itemize}
    \item 1) It has at least one detection with luminosity above the Eddington limit within the uncertainty ($\verb|Luminosity|+\verb|LuminosityErr|>10^{39}$\,\lum) in at least one of the potential host galaxies.
\end{itemize}

This check identified 5\,943 entries of diverse nature, corresponding to 3\,280 sources in 2\,729 galaxies. 2\,205 of these objects had been flagged as central, while only 856 objects in 552 galaxies were clean of counterparts. Furthermore, some of them present large uncertainties in their data. Therefore, we imposed additional quality conditions for a source to qualify as a \textit{ULX candidate}:

\begin{itemize}
    \item 2) It has no identified counterpart.
    \item 3) It holds $\verb|SC_Luminosity|>\verb|SC_LuminosityErr|$, as otherwise it would indicate that its source parameters are unreliable.
    \item 4) It has $\verb|SC_SUM_FLAG|<1$. This way, we only considered sources for which none of the individual detections was flagged as probably spurious.
    \item 5) It has a single potential host galaxy ($\verb|n_Galaxies|=1$), as otherwise their distances and luminosities are unreliable.
\end{itemize}

This left originally 730 ULX candidates in 490 galaxies. Additionally, we created a subcategory of \textit{bright} ULX candidates, which follows that $\verb|Luminosity|+\verb|LuminosityErr|>5\times10^{40}$\,\lum, into which only 130 in 122 galaxies were included. We also found 7 sources that, despite having more than one potential host galaxy ($\verb|n_Galaxies|>1$), their distances were differing only by a small fraction without affecting their ULX status. These sources were included to the ULX count, one of them being a bright candidate.

Sources that were below the ULX luminosity regime in all of their detections but still hold conditions 2) to 5) are simply referred to as \textit{quality sources}. Many of these sources and low-luminosity interlopers identified in \textit{GaiaDR2}, \textit{PanSTARRS1} and \textit{SIMBAD} are found in nearby galaxies, as showcased by Figure~\ref{M51}.

\subsection{Manual inspection of candidates and contaminants}\label{manual}

Our procedure revealed 50 ULX candidates with $L_\textrm{X}>10^{41}$\,\lum, well within the AGN luminosity regime \citep{Brandt2015}. However, we had to consider the possibility of some of them being contaminants in disguise that survived the filtering pipeline. With the aim to identify potential counterparts, we inspected the available optical and X-ray images from the \textit{PanSTARRS1} and \textit{XMM-Newton} surveys for 112 bright ULX candidates. Here we discuss some of the relevant results:

\begin{figure}[t]
\centering
\includegraphics[width=0.9\columnwidth]{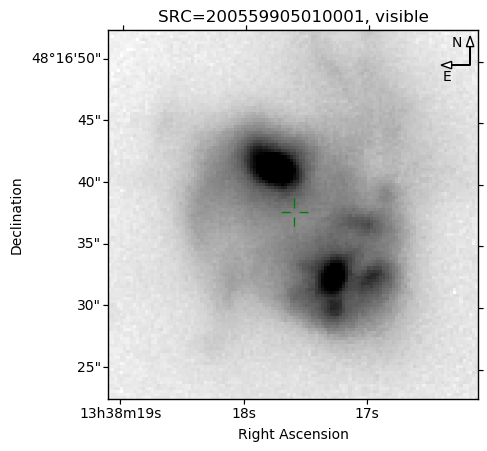}
\caption{\textit{PanSTARRS1} optical image of the location of source SRCID=200559905010001. The X-ray source (not shown here) is a blend of the two cores of the interacting pair.}
\label{200559905010001}
\end{figure}

\begin{figure}[t]
\centering
\includegraphics[width=0.9\columnwidth]{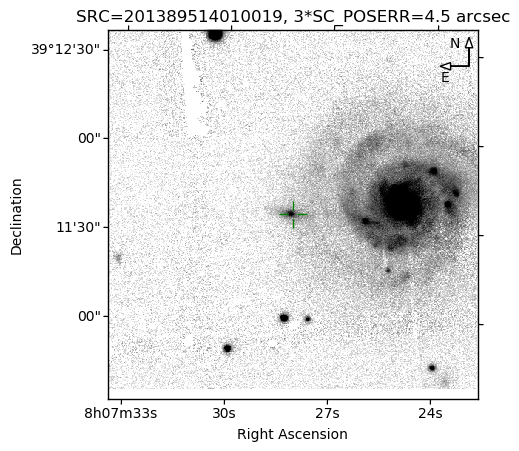}
\caption{\textit{PanSTARRS1} optical image of the location of source SRCID=201389514010019. The source lies very clearly on top of a background galaxy.}
\label{201389514010019}
\end{figure}

Source $\verb|SRCID|=200559905010001$ (Figure~\ref{200559905010001}), with detections of $L_\textrm{X}=(4.3\pm1.3)\times10^{42}$\,\lum\ and $L_\textrm{X}=(8.2\pm2.3)\times10^{42}$\,\lum, is most likely a blend of the two AGNs in the interacting galaxy pair \object{NGC 5256}.

Source $\verb|SRCID|=206701401010002$ presents $\langle L_\textrm{X}\rangle=(4.5\pm0.9)\times10^{41}$~\lum\,and is found in the interacting galaxy pair \object{II Zw}. This pair has been thoroughly studied in the X-rays \citep{Inami2010,Iwasawa2011}, and the \textit{XMM-Newton} source is most likely a blend of sources A, C and D from \cite{Goldader1997}.

Source $\verb|SRCID|=201389514010019$ (Figure~\ref{201389514010019}) clearly points towards a background galaxy, despite it being listed as belonging to galaxy \object{NGC 2528}. According to the \textit{NASA/IPAC Extragalactic Database}\footnote{\url{https://ned.ipac.caltech.edu/}} (\textit{NED}), this galaxy is at $z=0.13$.

A further amount of 8 sources were revealed as spurious detections around an over-saturated region on the X-ray image. All of the mentioned sources were flagged accordingly.

Unfortunately, 19 of the sources did not have \textit{PanSTARRS1} images available, so we inspected their positions on \textit{NED} to look for possible counterparts. This is particularly relevant for objects with ecliptic latitude below $-30^\circ$, as this area is not entirely covered by the \textit{PanSTARRS1} survey. 15 additional sources showed \textit{PanSTARRS1} counterparts of unclear nature, so we double checked them with \textit{NED} too. The results were the following:

Sources $\verb|SRCID|=200936502010001$ and $\verb|SRCID|=200029702010002$ were found to be the central sources of their host galaxies, \object{NGC 5128} and \object{UGC 01841}, due to a discrepancy in the value for the central coordinates with \textit{HECATE}. Source $\verb|SRCID|=201241101010001$ was identified as the QSO \object{MR 0205}. Finally, four remaining sources were also identified with distant background galaxies.

There is another side to the manual inspection of sources. As ULXs are more commonly associated with star-forming regions \citep{Kaaret2017}, some genuine ULXs are located within or next to optically bright H II regions of spiral and irregular galaxies. This caused confusion in the filtering step involving \textit{PanSTARRS1}, and otherwise good candidates were flagged as interlopers. Therefore, we performed manual inspection for 128 \textit{PanSTARRS1} associations that held $\verb|SC_SUM_FLAG|\leq1$, $\verb|SC_Luminosity|+\verb|SC_LuminosityErr|\geq10^{39}$\,\lum\ and $\verb|SC_Luminosity|>\verb|SC_LuminosityErr|$ with the intention of recovering genuine ULX candidates.

\begin{figure}[t!]
\centering
\includegraphics[width=0.9\columnwidth]{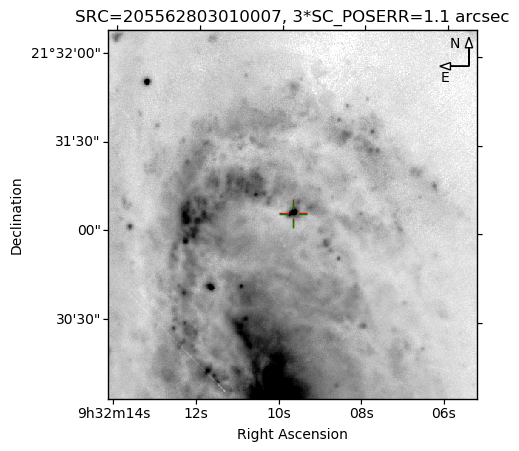}
\caption{\textit{PanSTARRS1} optical image of the location of source SRCID=205562803010007 (green marker) in galaxy \object{NGC 2903}. The red marker indicates the position of the \textit{PanSTARRS1} counterpart identified by the pipeline. The optical counterpart constitutes an H II region.}
\label{205562803010007}
\end{figure}

72 sources were matched to bright nebulous features in spiral galaxies (see Figure~\ref{205562803010007} for an example). \textit{NED} based inspection of 10 further sources also suggested H II region nature or having only X-ray counterparts, and therefore were re-flagged as clean. Source $\verb|SRCID|=201109302010027$ in particular was not matched to any potential contaminant in \textit{NED}, but it showed two possible X-ray counterparts. It was kept as an interloper due to its possibly blended nature.

Another set of 11 sources were identified with infrared sources also in \textit{NED}, while four more sources where found to coincide with background galaxies. These sources were kept in the group of interlopers.

Finally, we searched for possible extragalactic objects in \textit{GaiaDR2}. \cite{Shu2019} draw 3\,175\,537 AGN candidates from 641\,266\,363 \textit{GaiaDR2} sources, a fraction of 0.5\%, which suggested further possible undetected extragalactic interlopers in our catalogue. Therefore, we performed a final cross-match of the remaining unmatched sources with \textit{GaiaDR2}, this time applying condition (\ref{optical objects}) to decide whether to flag the object as a possible interloper. 23 ULX were matched with a \textit{GaiaDR2} source. We inspected all of them manually in the \textit{NED} database and found that 14 of them did not have a known counterpart, so they were re-flagged as clean. In contrast, 7 were found to have infrared counterparts, $\verb|SRCID|=203049401010029$ was matched to a UV source and $\verb|SRCID|=207843705010011$ was found to be the central source of \object{NGC 7632}, and therefore they were accordingly confirmed as interlopers.

The final version of the catalogue after the manual inspections yielded 1\,452 detections of 779 ULX candidates in 517 galaxies, and 163 detections of 94 bright ULX candidates of quality in 94 galaxies. 

\subsection{Luminosity and complete sub-samples}\label{sub-samples}

As low-luminosity sources become increasingly harder to detect with increasing distance, our catalogue presents a bias towards brighter sources at large distances, hampering any potential study of the ULX population properties. Following the methodology of \cite{Earnshaw2019}, we created the luminosity sub-samples to ensure the completeness of the set of sources above a luminosity threshold $L_\textrm{min}$ inside of a radius $D_\textrm{max}$.

We established a minimum flux for a source to be detected, $F_\textrm{min}$, from which we compute the maximum distances $D_\textrm{max}$ at which sources with luminosity $L_\textrm{min}$ can be seen. Then, we selected all objects with $D<D_\textrm{max}$ in each case to make sure that every source with $L>L_\textrm{min}$ is accounted for, mitigating this way the bias towards brighter sources in the $L>L_\textrm{min}$ regime. From these, we then selected all quality sources and ULX candidates whose host galaxy's isophotal dimensions are known ($\verb|MATCH_FLAG|=0$) and that lie at least $25^\circ$ away from the galactic plane. This last condition ensures minimal photoelectric absorption from the Milky Way.

\begin{figure}[t]
\centering
\includegraphics[width=0.9\columnwidth]{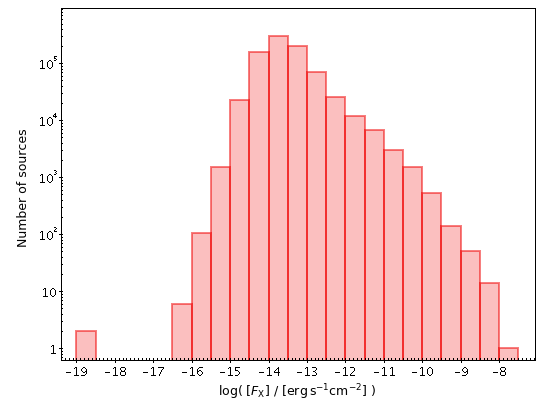}
\caption{Distribution of detection fluxes within \textit{4XMM-DR9}, regardless of their nature. Most detections (77\%) have been detected with fluxes higher than $F_\textrm{X}\approx10^{-14}$\fint.}
\label{fluxDist}
\end{figure}

\begin{figure}[t]
\centering
\includegraphics[width=0.9\columnwidth]{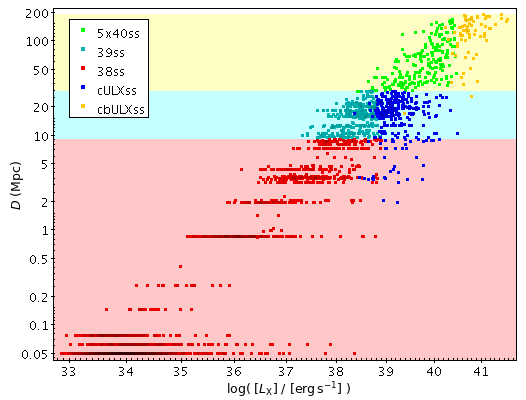}
\caption{Distance-luminosity dispersion of all quality sources within the the \textbf{38ss}, \textbf{39ss} and \textbf{5x40ss}, and the \textbf{cULXss} and the \textbf{cbULXss}, with an imposed sensitivity of $10^{-14}$~\fint. The distance cuts correspond to $D_\textrm{max}\approx9$, 29 and 204~Mpc.}
\label{CompleteSubsamples}
\end{figure}

The most controversial aspect of this method is the choice of $F_\textrm{min}$, since there is no hard detection threshold for the \textit{XMM-Newton} observatory as it depends on exposure time and background intensity. To work around this, we observed in Figure~\ref{fluxDist} that the distribution of all detection fluxes in \textit{4XMM-DR9} peaks early at $F_\textrm{X}\approx10^{-14}$\,\fint mark, with 77\% of the detections at higher fluxes, so we adopted this value as the baseline for \textit{XMM-Newton}'s sensitivity.

With this in hand, we built the luminosity sub-samples \textbf{38ss}, \textbf{39ss} and \textbf{5x40ss}, with corresponding $L_\textrm{min}=10^{38}$, $10^{39}$ and $5\times10^{40}$\,\lum\ and $D_\textrm{max}\approx9$, 29 and 204\,Mpc. We then selected all ULX candidates with $\verb|n_Galaxies|=1$ belonging to \textbf{39ss} to build the complete ULX sub-sample (\textbf{cULXss}), and we did the same with the bright ULX candidates and the \textbf{5x40ss}, building the complete bright ULX sub-sample (\textbf{cbULXss}).

The effectiveness of this method is well illustrated in Figure~\ref{CompleteSubsamples}, where the bias towards brighter sources has been significantly downplayed by excluding sources with $D>D_\textrm{max}$ from the sub-sample. As shown in Table~\ref{samples}, these two sub-samples contain a reduced amount of 292 and 69 sources. The sub-samples are also helpful in an additional way: due to the clustering nature of ULXs, objects at $D>10$~Mpc are prone to blending due to the limited angular resolution \textit{XMM-Newton}. The cut at $D_\textrm{max}=29$~Mpc also helps in the mitigation of this bias.

\section{Results}\label{resultsSection}

\subsection{Contents of the catalogue}\label{contents}

\begin{table}[t]
 \cprotect\caption[]{\label{samples} Number of detections, sources, host galaxies, mean logarithm of luminosity and median distance of each population sample with $\verb|n_Galaxies|=1$. All objects with multiple possible host galaxies, including seven ULX candidates, have been excluded from this table.}
 \centering
\begin{tabular}{lccccc}
 \hline
 \hline
 \multirow{2}{*}{Sample} & \multirow{2}{*}{Dets.} & \multirow{2}{*}{Srcs.} & \multirow{2}{*}{Gals.} & \multirow{2}{*}{\small{$\big\langle\log\left(\frac{L_\textrm{X}}{\textrm{\lum}}\right)\big\rangle$}} & $\left[ D \right]$ \\
  & & & & & (Mpc) \\
\hline
 Quality          & 18\,421 & 6\,385 & 607 & 36.1 & 0.77 \\
 ULX              & 1\,439  & 772  & 515 & 39.6 & 28.4 \\
 B. ULX           & 162   & 93   & 93  & 40.8 & 127  \\
 \textbf{38ss}    & 9\,594  & 3\,290 & 56  & 34.8 & 0.05 \\
 \textbf{39ss}    & 10\,567 & 3\,864 & 209 & 35.3 & 0.05 \\
 \textbf{5x40ss}  & 10\,959 & 4\,148 & 452 & 35.7 & 0.06 \\
 \textbf{cULXss}  & 680   & 292  & 146  & 39.2 & 17.5 \\
 \textbf{cbULXss} & 126   & 69   & 69  & 40.8 & 128  \\
\hline
\end{tabular}
\end{table}

\begin{table}[t]
 \cprotect\caption[]{\label{results} List with the number of detections, sources and galaxies with every kind of \verb|CONT_FLAG| and $\verb|n_Galaxies|=1$. \verb|CONT_FLAG| is the flag indicating whether a source is clean (\textit{``none''}) or the catalogue or step where a counterpart has been identified.}
 \centering
\begin{tabular}{lccc}
 \hline
 \hline
 $\textrm{CONT\_FLAG}$ & Dets. & Srcs. & Gals. \\
 \hline
 Any                             & 49\,816 & 23\,120 & 2\,700 \\
 \textit{``none''}               & 23\,340 & 7\,657  & 607  \\
 \textit{``none(PanSTARRS1)''}   & 75    & 63    & 60   \\
 \textit{``none(NED)''}          & 36    & 16    & 16    \\
 \textit{``central''}            & 3\,514  & 2\,173  & 2\,170 \\
 \textit{``GaiaDR2''}            & 13\,402 & 9\,666   & 88   \\
 \textit{``Tycho2''}             & 96    & 11    & 6    \\
 \textit{``SDSS$\_$DR14''}       & 140   & 33    & 24   \\
 \textit{``VeronQSO''}           & 135   & 38    & 11   \\
 \textit{``SIMBAD''}             & 5\,745  & 2\,224  & 65   \\
 \textit{``PanSTARRS1''}         & 2\,980  & 1\,119  & 136  \\
 \textit{``manual(PanSTARRS1)''} & 10    & 9     & 3    \\
 \textit{``manual(NED)''}        & 38    & 26    & 25   \\
\hline
\end{tabular}
\end{table}

Tables \ref{samples} and \ref{results} summarise the content of our final catalogue. It contains 50\,446 entries corresponding to 23\,262 sources, out of which 23\,120 are associated with a single galaxy ($\verb|n_Galaxies|=1$). X-ray sources with \textit{GaiaDR2} associations form the largest group of contaminants, containing a 41.8\% of the total number of sources. This is due to the 8\,437 matches with objects of large optical to X-ray flux ratio, 8\,270 of which are located in the Magellanic Clouds and M31, indicating a possible stellar nature. Nonetheless, none of them fall within the ULX X-ray luminosity regime so no further regard is paid to them. Central objects follow it, constituting a 9.7\% of the catalogue. They are followed by a 8.4\% of objects directly identified as AGNs or QSOs in \textit{SDSS\_DR14}, \textit{VeronQSO} and \textit{SIMBAD}. A further 7.1\% have been identified as stars in \textit{GaiaDR2}, \textit{Tycho2} or \textit{SIMBAD}. Only 4.9\% of the objects have been selected as potential interlopers from the \textit{PanSTARRS1} survey, and a residual fraction is constituted by supernovae, infrared sources, background galaxies and a single ultraviolet source.

A total of 3\,274 (14.1\%) objects have been detected at least once within the ULX regime ($L_\textrm{X}+\Delta L_\textrm{X}>10^{39}$\,\lum). 2\,208 of these consist of central sources, and only 779 (3.3\%) qualify as ULX candidates according to the thresholds established in Sections~\ref{ULXcandidates}~and~\ref{manual}, and 287 are interlopers of various kind. 516 of the ULX candidates have been detected at least once within the ULX regime with high certainty ($L_\textrm{X}-\Delta L_\textrm{X}>10^{39}$\,\lum), and 666 with high likelihood ($L_\textrm{X}>10^{39}$\,\lum). 761 have at least a detection well above the NS Eddington limit ($L_\textrm{X}-\Delta L_\textrm{X}>1.8\times10^{38}$\,\lum). A sub-set of 94 also qualify as bright ULX candidates ($L_\textrm{X}+\Delta L_\textrm{X}>5\times10^{40}$\,\lum), and all of them are the only candidate in their host galaxy. Among some of the candidates we find well-known objects such as the NS-ULXs \object{NGC 7793 P13}, \object{NGC 5907 ULX-1}, the IMBH candidate \object{NGC 2276-3c}, and \object{M51-ULS-1}, host of the extragalactic exoplanet candidate M51-ULS-1b \citep{DiStefano2020}. Other known objects such as the IMBH candidates \object{M82 X-1} and \object{ESO 243-49 HLX-1} also appear in the catalogue, but with $\verb|SC_SUM_FLAG|>1$.

Only sources and candidates with $\verb|n_Galaxies|=1$ are considered during the remainder of this section due to their more reliable luminosities and host galaxy associations.

\subsection{ULX distribution and abundances}

\begin{figure}[t]
\centering
\includegraphics[width=0.9\columnwidth]{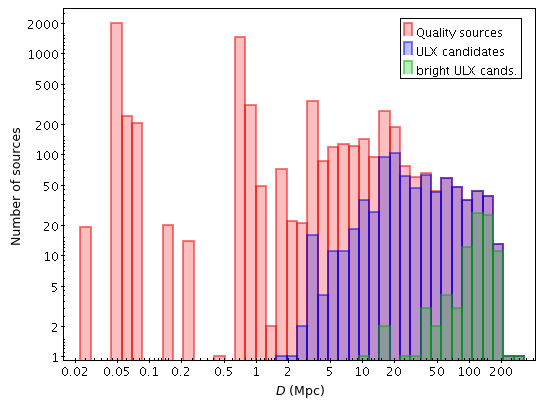}
\caption{Distance distribution of all sources of quality, ULX candidates and bright ULX candidates in our catalogue.}
\label{distances}
\end{figure}

In Figure~\ref{distances}, it can be seen that all ULX candidates are found at $D>1$~Mpc, while bright candidates are found at $D\gtrsim 20$~Mpc. Figure~\ref{distances} also illustrates how at $D\gtrsim 100$~Mpc the bright candidates dominate almost completely due to the bias in favor of brighter sources as discussed in Section~\ref{sub-samples} and possibly source blending. This is consistent with the information in Table~\ref{samples}, which shows that median distances of ULX and bright ULX candidates are 28.4~Mpc and 127~Mpc. Sources of quality cluster in the nearest galaxies for the same reason. They constitute a population of low luminosity X-ray sources the dominate the serendipitous X-ray sky in the local group of galaxies. For instance, M31, M33 and the Magellanic Clouds alone already concentrate a 17\% of all sources of quality. The X-ray content of the local group is not directly relevant to the ULX population, as it is mostly composed of typical X-ray binaries and supernovae remnants and it has already been thoroughly studied \citep[e.g.,][]{Sturm2013,Pietsch2008}. Therefore, during the raeminder of this paper we focus most of our attention to sources hosted by galaxies at $D>1$~Mpc.

\begin{figure}[t]
\centering
\includegraphics[width=0.9\columnwidth]{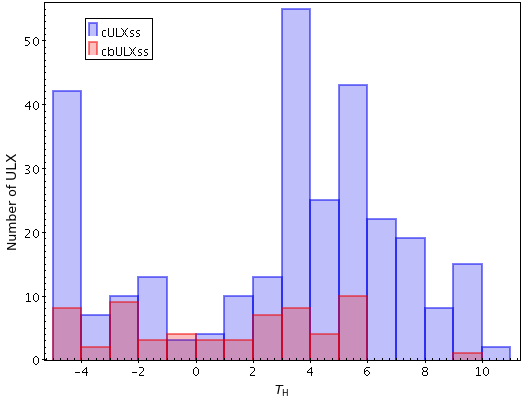}
\caption{Hubble type distributions of the \textbf{cULXss} and \textbf{cbULXss}. In both cases, there is a spike of ULXs in the earliest galaxies.}
\label{HubbleDist}
\end{figure}

\begin{figure}[t!]
\centering
\includegraphics[width=\columnwidth]{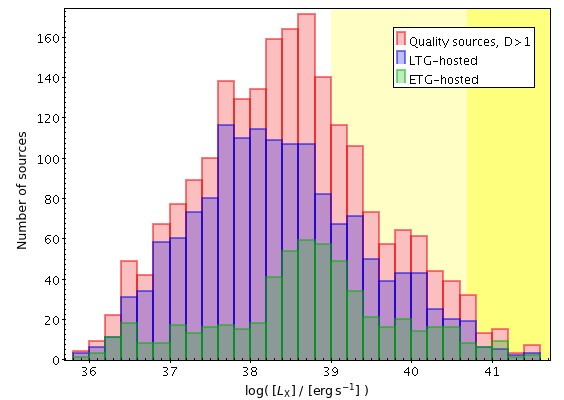}
\caption{Luminosity distribution of the sources of quality and ULX candidates in our catalogue divided by the morphological types of galaxies. A distance cut of $D>1$~Mpc has been imposed to avoid contamination from the Local Group, heavily dominated by low-luminosity sources.}
\label{MorphDist}
\end{figure}

\begin{figure}[t]
\centering
\includegraphics[width=0.9\columnwidth]{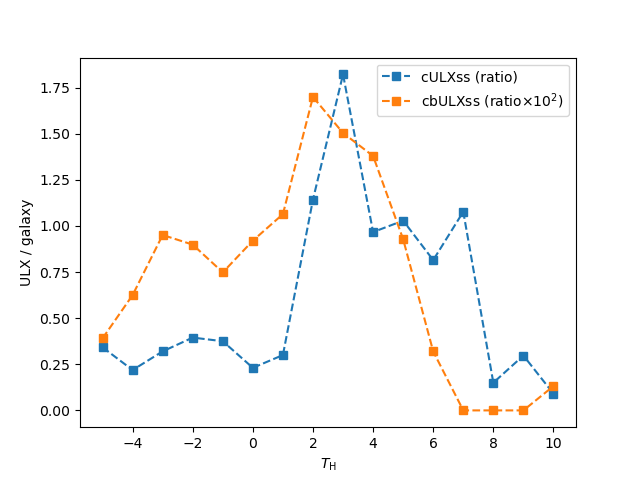}
\caption{Distribution of ULX frequencies (ULX/galaxy) in the \textbf{cULXss} and \textbf{cbULXss} across the morphological sub-types of host galaxies. The ratios from the \textbf{cbULXss} have been multiplied by 100 to aid visibility.}
\label{ULXfreq}
\end{figure}

\begin{table}[h]
 \cprotect\caption[]{\label{GalaxyULXs} Number of ULX candidates in the \textbf{cULXss} and the \textbf{cbULXss}, number of galaxies that host them, number of galaxies that could potentially hosted them, ULX frequencies and fraction of galaxies with candidates, divided according to morphological groups of their host galaxies. No distinction is made between bright ULXs and their host galaxies as no more than one candidate has been found in each.}
 \centering
\begin{tabular}{lcccccc}
 \hline
 \hline
 Sample &\multicolumn{6}{c}{\textbf{cULXss}} \\
 \small{Subsample}    & All   & El.  & Le.  & ESp. & LSp. & Ir.   \\
 \hline
 \small{ULXs}         & 292   & 49   & 26   & 107    & 92     & 17    \\
 \small{Galaxies}     & 149   & 18   & 16   & 53     & 45     & 13    \\
 \small{Pot. gals.}   & 722   & 151  & 75   & 109    & 114    & 147   \\
 \small{ULX / gal.}   & 0.40  & 0.32 & 0.35 & 0.98   & 0.81   & 0.12  \\
 \small{Gal. frac.} & 0.21  & 0.12 & 0.21 & 0.49   & 0.37   & 0.09  \\
 \hline
 \hline
 Sample &\multicolumn{6}{c}{\textbf{cbULXss}} \\
 \small{Subsample}        & All   & El.  & Le.  & ESp. & LSp. & Ir.   \\
 \hline
 \small{ULXs}           & 69    & 10   & 16   & 25     & 10     & 1     \\
 \small{Pot. gals.}       & \small{12\,166} & \small{2\,174} & \small{1\,787} & \small{2\,052}   & \small{1\,416}   & 850   \\
 \tiny{ULX/gal.(\%)}      & 0.57  & 0.46 & 0.90 & 1.22   & 0.71   & 0.12  \\
 \hline
\end{tabular}
\end{table}

\begin{figure}[t]
\centering
\includegraphics[width=0.9\columnwidth]{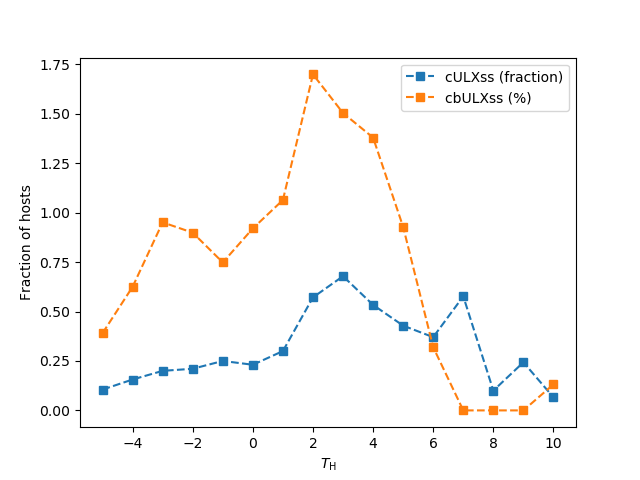}
\caption{Distribution of fraction of host galaxies in the \textbf{cULXss} and \textbf{cbULXss} across their morphological sub-types. The values from \textbf{cbULXss} are shown in \% to aid visibility.}
\label{HostFreq}
\end{figure}

In Table~\ref{GalaxyULXs}, it can be seen that objects from the \textbf{cULXss} are most commonly associated with early spiral galaxies (ESp.,~$0\leq T_\textrm{H}<5$) and late spiral galaxies (LSp.,~$5\leq T_\textrm{H}<9$). However, while 68.1\% of the \textbf{cULXss} candidates are found in spiral galaxies, only for 50.7\% of the objects in \textbf{cbULXss} dataset holds the same trend. In Figure~\ref{HubbleDist} it can be seen how the overwhelming majority of \textbf{cULXss} objects concentrate in spiral galaxies in the range of $0\leq T_\textrm{H}<10$, with a spike in elliptical galaxies (El.,~$T_\textrm{H}<-3$). For \textbf{cbULXss}, the distribution is similar, but the spike in elliptical galaxies is more prominent. In both samples, lenticular galaxies (Le.,~$-3\leq T_\textrm{H}<0$) and irregular galaxies (Ir.,~$T_\textrm{H}\geq9$) are unpreferred. Figure~\ref{MorphDist} also showcases that while ULX candidates of quality are more frequently found in late-type galaxies (LTG, $T_\textrm{H}\geq0$) than in early-type galaxies (ETG, $T_\textrm{H}<0$), this distinction becomes less clear for the bright ULX candidates.

Besides looking how ULX candidates distribute themselves among galaxy morphological types, it is also interesting to investigate the fraction of galaxies that contain any ULX and general ULX frequencies in them. For this, we build the set of all galaxies in the field of view that would be included in the \textbf{cULXss} or \textbf{cbULXss} if hosting a ULX was not a requirement candidate, with the extra condition of having $R1>3$" to exclude galaxies that can only contain central sources. In essence, this consists of the baseline of galaxies that would belong to the complete sub-samples if they host at least one ULX candidate, including the ones that actually do. This is shown in Table~\ref{GalaxyULXs} as the line of potential galaxies, and it is used implicitly in Table~\ref{GalaxyULXsSFR} to compute the mean SFR and $M_*$ values. In Table~\ref{GalaxyULXs}, it is seen that 21\% of galaxies in the \textbf{cULXss} contains at least a ULX candidate. For elliptical galaxies, this gets reduced to 12\%, for lenticular galaxies it is 21\%, for early spirals 49\%, and for late spirals 37\%. The general ULX rate is $\sim$0.4 ULX/galaxy, being the highest at $\sim$0.98 ULX/galaxy in early spiral galaxies. Figures~\ref{ULXfreq}~and~\ref{HostFreq} show this results in more detail for the different galaxy morphological sub-types, with ULX frequencies peaking at $\sim$1.8 ULX/galaxy for $T_\textrm{H}\sim3$ galaxies, of which $\sim$70\% host ULX candidates. These results are mostly in agreement with earlier works, that either found a larger fraction of ULXs in late-type galaxies \citep{Swartz2011} or find higher abundances in late-type galaxies \citep{Earnshaw2019,Kovlakas2020}. However, \cite{Kovlakas2020} find that the galaxies with a higher chance of hosting ULXs are $T_\textrm{H}\sim 5$ spirals instead.

\begin{table}[h]
 \cprotect\caption[]{\label{GalaxyULXsSFR} Mean star-formation rate, stellar mass, and derived ULX rates for all morphological types. Computed from all potential galaxies from which we have SFR and $M_*$ information. As stated in Section~\ref{GraySample}, SFR information is only taken into account for late-type galaxies.}
 \centering
\begin{tabular}{lccccc}
 \hline
 \hline
 Sample &\multicolumn{5}{c}{\textbf{cULXss}} \\
 \small{Subsample} & El. & Le.  & ESp. & LSp. & Ir.   \\
 \hline
 $\langle\textrm{SFR}\rangle$ & \multirow{2}{*}{...} & \multirow{2}{*}{...} & \multirow{2}{*}{1.99}   & \multirow{2}{*}{0.80}   & \multirow{2}{*}{0.02}  \\
 \small{(\sfr)} & & & & & \\
 $\langle M_*\rangle$ & \multirow{2}{*}{2.33} & \multirow{2}{*}{2.59} & \multirow{2}{*}{4.47} & \multirow{2}{*}{1.08}   & \multirow{2}{*}{0.14}  \\
 \small{($10^{10}\,\textrm{M}_\odot$)} & & & & & \\
 $\langle \textrm{sSFR} \rangle$  & \multirow{2}{*}{...} & \multirow{2}{*}{...} & \multirow{2}{*}{4.46}   & \multirow{2}{*}{7.37}   & \multirow{2}{*}{12.5}  \\
  \small{($10^{-11}\textrm{yr}^{-1}$)} & & & & & \\
 $\langle\textrm{ULX / SFR}\rangle$   & \multirow{2}{*}{...} & \multirow{2}{*}{...} & \multirow{2}{*}{0.49}   & \multirow{2}{*}{1.01}   & \multirow{2}{*}{6.41}  \\
  \small{(\sfr)$^{-1}$} & & & & & \\ 
 $\langle \textrm{ULX / }M_*\rangle$  & \multirow{2}{*}{1.39} & \multirow{2}{*}{1.34} & \multirow{2}{*}{2.20}   & \multirow{2}{*}{7.48}   & \multirow{2}{*}{80.2}  \\
 \small{($10^{-11}\,\textrm{M}_\odot^{-1}$)} & & & & & \\ 
 \hline
 \hline
 Sample &\multicolumn{5}{c}{\textbf{cbULXss}} \\
 \small{Subsample} & El. & Le.  & ESp. & LSp. & Ir.   \\
 \hline
 $\langle\textrm{SFR}\rangle$   & \multirow{2}{*}{...} & \multirow{2}{*}{...} & \multirow{2}{*}{3.25}   & \multirow{2}{*}{1.29}   & \multirow{2}{*}{1.14}  \\
  \small{(\sfr)} & & & & & \\
 $\langle M_*\rangle$   & \multirow{2}{*}{5.26} & \multirow{2}{*}{5.63} & \multirow{2}{*}{5.31}   & \multirow{2}{*}{1.32}   & \multirow{2}{*}{2.73}  \\
  \small{($10^{10}\,\textrm{M}_\odot$)} & & & & & \\
 $\langle \textrm{sSFR} \rangle$   & ... & ... & 6.12 & 9.80 & 4.19  \\
  \small{($10^{-11}\textrm{yr}^{-1}$)} & & & & & \\
 $\langle\textrm{ULX / SFR}\rangle$  & \multirow{2}{*}{...} & \multirow{2}{*}{...} & \multirow{2}{*}{3.75} & \multirow{2}{*}{5.46}   & \multirow{2}{*}{1.03}  \\
  \small{$10^{-3}$(\sfr)$^{-1}$} & & & & & \\ 
 $\langle \textrm{ULX / }M_*\rangle$   & \multirow{2}{*}{0.87} & \multirow{2}{*}{1.59} & \multirow{2}{*}{2.29} & \multirow{2}{*}{5.34} & \multirow{2}{*}{0.43} \\
 \small{($10^{-13}\,\textrm{M}_\odot^{-1}$)} & & & & & \\ 
 \hline
\end{tabular}
\end{table}

\begin{figure}[t!]
\centering
\includegraphics[width=0.9\columnwidth]{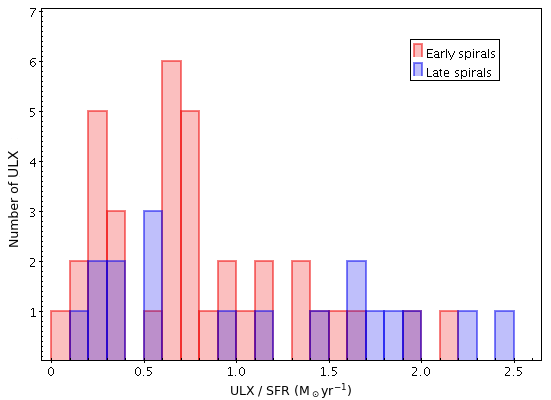}
\caption{Histogram of the number of ULX per star-formation rate for early and late spiral galaxies from the \textbf{cULXss}. Extremely large numbers on the right are due to lucky galaxies with low SFR that host a single ULX have been cut out of the picture.}
\label{ULXperSFR}
\end{figure}

To find an explanation for these distributions, we need to explore the properties of host galaxies, and more particularly, the relationship between ULX abundances, host SFR and $M_*$. In spiral galaxies with high specific-SFR ($\textrm{sSFR}=\textrm{SFR}/M_*$) values and that are dominated by young stellar populations, the ULX population is typically associated with HMXB evolution timescales of $\tau\sim$100~Myr \citep{Wiktorowicz2017}, the rates of which are expected to scale with the SFR. In Table~\ref{GalaxyULXsSFR}, we show how spiral galaxies are typically the ones with the highest SFR as given in \textit{HECATE}. Early spirals have higher SFR absolute values than late spirals ($\sim$1.99 vs. $\sim$0.80~\sfr), but as expected they also have lower sSFR values. Despite ULXs being more abundant in early spirals, the ULX frequency per SFR is higher in late spirals ($\sim$0.49 vs. $\sim$1.01~\psfr). The same holds for the number of ULXs per unit mass, and the whole picture is exagerated for irregular galaxies. Figure~\ref{ULXperSFR} also shows the distribution of individual ULX/SFR galaxy rates for early and late spiral galaxies in the \textbf{cULXss} that contain available SFR information. The distributions peak at $\sim$0.65 and $\sim$0.55~ULX\psfr each. \cite{Wiktorowicz2017} predict the existence of 400 ULXs in a galaxy with solar metallicity, $M_*=6\times10^{10}$~M$_\odot$ and $\textrm{SFR}=600$~\sfr\ for a period of 100~Myr, which implies 0.67~ULX\psfr, staying well withing the range of values provided by our sample.

\begin{figure}[t!]
\centering
\includegraphics[width=0.9\columnwidth]{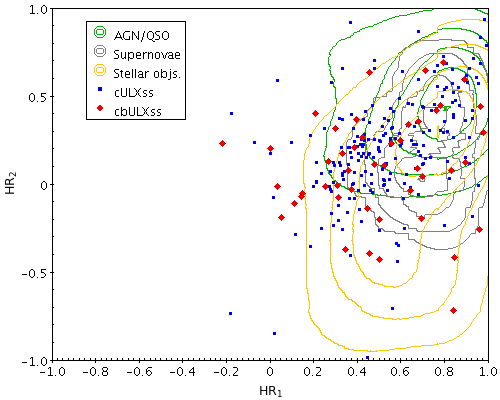}
\caption{Hardness ratio dispersion of the \textbf{cULXss} and \textbf{cbULXss} against the abundance contours of some populations of interlopers. AGNs and QSO include both confirmed and candidates. Only hardness ratios with uncertainties lower than 0.2 are accepted for the sake of reliability.}
\label{HR_AGN_QSO}
\end{figure}

\begin{figure}[t!]
\centering
\includegraphics[width=0.9\columnwidth]{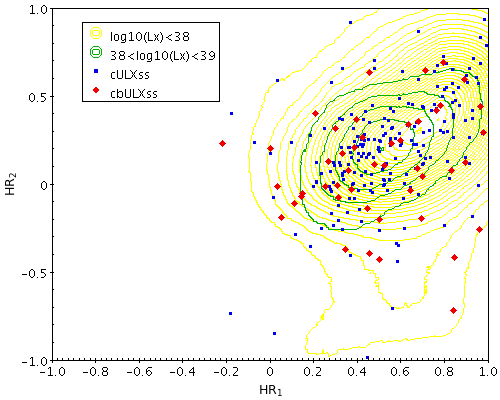}
\caption{Hardness ratio dispersion of the \textbf{cULXss} and \textbf{cbULXss} against the abundance contours of lower luminosity objects present in the \textbf{38ss}. Only hardness ratios with uncertainties lower than 0.2 are accepted for the sake of reliability.}
\label{HRpop}
\end{figure}

For elliptical and lenticular galaxies, different considerations need to be applied. As expressed by \cite{Wiktorowicz2017}, there is a ULX subpopulation constituted by LMXBs that reach Roche-lobe overflow at $\tau\sim$1~Gyr instead. This evolutionary path explains the presence of ULXs in galaxies with very low star-forming activity. In this case, ULX frequencies are not expected to depend significantly on their current SFR values.

On the other hand, objects in the \textbf{cbULXss} provide a slightly different picture. All of them are located only beyond the $D\gtrsim20$\,Mpc mark and have extremely low abundances, as shown in Table~\ref{GalaxyULXs}. Additionally, all of them are the only bright ULX candidate in their host galaxies. From \cite{Wiktorowicz2017} predictions the rate of objects with $L_\textbf{X}>10^{40}$\,\lum, we estimate a typical number of 0.023~bright ULX\psfr\ in star-forming galaxies. However, we only measure 0.0037~bright ULX\psfr\ and 0.0055~bright ULX\psfr\ for early and late-spiral galaxies, which is an order of magnitude smaller. This can be partially explained by considering that our luminosity cut of $L_\textbf{X}>5\times10^{40}$\,\lum\ is slightly higher, and that the ULX luminosity distribution in general decays exponentially in most studies. Nonetheless, a total number of only 36 objects are involved in these estimations. A larger sample will need to be used in the future to extract more reliable conclusions.

\subsection{Spectral properties of the catalogue}\label{spectral}

The X-ray spectral properties of the identified sources can provide additional information regarding their nature \citep[e.g.,][]{Earnshaw2019,Walton2011}. Therefore, we proceed with the inspection of the spectral properties of the objects within our catalogue. We use the hardness ratios, defined as
\begin{equation}\textrm{HR}_i=\frac{R_{i+1}-R_i}{R_{i+1}+R_i}\textrm{,}\end{equation}
where $R_i$ and $R_{i+1}$ are the count rates in the consecutive energy bands $i$ and $i+1$. With this quantity, spectral hardness can be compared between different sources, with larger values indicating harder sources. The \textit{4XMM-DR9} computes the source $HR_i$ values from the count rates in the 0.2--0.5; 0.5--1.0; 1.0--2.0; 2.0--4.5; and 4.5--12.0~keV energy bands, both from the PN and MOS cameras\footnote{\url{http://xmmssc.irap.omp.eu/}}, $HR_1$ and $HR_2$ being the most useful thanks to larger photon counts in lower energy bands.

\begin{table}[t]
 \cprotect\caption[]{\label{HRcont} Number of objects in every contaminant population and their mean hardness ratios. Contaminants found in the \textit{PanSTARRS1} survey and extragalactic objects from \textit{PanSTARRS1} are not included due to the diverse nature of the matched objects.}
 \centering
\begin{tabular}{lccccccc}
\hline
\hline
 OBJ$\_$TYPE & \# & $\langle \textrm{HR}_\textrm{1}\rangle$ & $\langle \textrm{HR}_\textrm{2}\rangle$ & $\langle \textrm{HR}_\textrm{3}\rangle$ & $\langle \textrm{HR}_\textrm{4}\rangle$ \\
\hline
 \textit{``central''}      & 2\,246 & 0.52 & $-$0.10 & $-$0.34 & $-$0.18 \\
 Stellar objs.             & 1\,652 & 0.45 & $-$0.05 & $-$0.47 & $-$0.10 \\
 \textit{``AGN''}          & 320    & 0.61 &    0.43 & $-$0.21 & $-$0.26 \\
 \textit{``AGN$\_$Cand.''} & 1\,416 & 0.59 &    0.52 & $-$0.20 & $-$0.23 \\
 \textit{``QSO''}          & 126    & 0.47 &    0.24 & $-$0.26 & $-$0.29 \\
 \textit{``QSO$\_$Cand.''} & 81     & 0.55 &    0.31 & $-$0.30 & $-$0.25 \\
 \textit{``SN''}           & 27     & 0.66 &    0.07 & $-$0.28 & $-$0.28 \\
 \textit{``IrS''}          & 17     & 0.42 &    0.13 & $-$0.53 & $-$0.20 \\
 \textit{``back. galaxy''} & 9      & 0.41 & $-$0.08 & $-$0.01 & $-$0.32 \\
\hline
\end{tabular}
\end{table}

\begin{table}[t!]
 \caption[]{\label{GalaxyHR} Mean hardness ratios of the \textbf{cULXss} and \textbf{cbULXss} populations, broken down according to the morphological type of host galaxies.}
 \centering
\begin{tabular}{lccccc}
 \hline
 \hline
 Population & Size & $\langle\textrm{HR}_1\rangle$ & $\langle\textrm{HR}_2\rangle$ & $\langle\textrm{HR}_3\rangle$ & $\langle\textrm{HR}_4\rangle$ \\
 \hline
 \textbf{cULXss}  & 292 & 0.51 & 0.24 & $-$0.25 & $-$0.50 \\
 El.-hosted       & 49  & 0.42 & 0.09 & $-$0.27 & $-$0.46 \\
 Le.-hosted       & 27  & 0.36 & 0.13 & $-$0.24 & $-$0.40 \\
 ESp.-hosted    & 107 & 0.49 & 0.24 & $-$0.25 & $-$0.52 \\
 LSp.-hosted    & 92  & 0.61 & 0.36 & $-$0.24 & $-$0.53 \\
 Ir.-hosted       & 17  & 0.53 & 0.21 & $-$0.35 & $-$0.52 \\
\hline
 \textbf{cbULXss} & 69  & 0.43 & 0.19 & $-$0.32 & $-$0.23 \\
 El.-hosted       & 10  & 0.41 & 0.06 & $-$0.49 & $-$0.39 \\
 Le.-hosted       & 16  & 0.43 & 0.10 & $-$0.33 & $-$0.24 \\
 ESp.-hosted    & 25  & 0.47 & 0.23 & $-$0.24 & $-$0.19 \\
 LSp.-hosted    & 10  & 0.43 & 0.35 & $-$0.21 & $-$0.22 \\
 Ir.-hosted       & 1   & 0.98 & 0.29 & $-$0.30 &  0.34 \\
\hline
\end{tabular}
\end{table}

From Table~\ref{HRcont}, it is apparent how the contaminant populations have distinct average spectral properties from each other. Supernovae, AGNs and QSOs tend to be the objects with the hardest spectra. Comparing the interloper hardness ratios with the ULX hardness ratios shown in Table~\ref{GalaxyHR} also shows that, as expected, the typical spectra of the ULX population is also distinct from those of stellar objects, supernovae, infrared sources and nuclear sources, while bearing the closest spectral resemblance with the AGN and QSO population. These properties are made more apparent by looking at Figure~\ref{HR_AGN_QSO}, where it can be seen that AGNs and QSOs cluster around slightly harder spectra than the \textbf{cULXss} and \textbf{cbULXss} samples, and that there is even less overlap of the ULX population with the supernovae and the stellar populations. Figure~\ref{HRpop} shows that the ULX population has the largest resemblance with the population of X-ray objects just below the ULX luminosity threshold ($10^{38}<L_\textrm{X}<10^{39}$\,\lum).

\begin{figure}[t!]
\centering
\includegraphics[width=0.9\columnwidth]{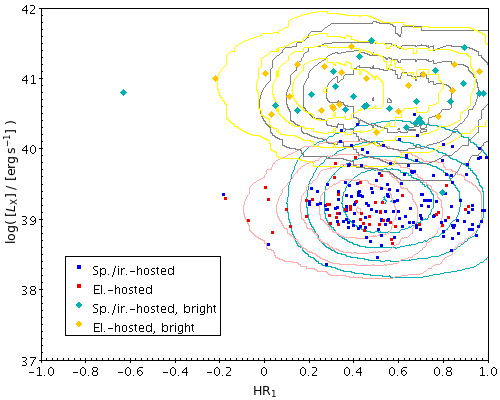}
\caption{Hardness ratio versus luminosity dispersion of objects in the \textbf{cULXss} (squares) and \textbf{cbULXss} (diamonds), distinguishing between objects hosted in late type galaxies ($T_\textrm{H}\geq0$) and early type galaxies ($T_\textrm{H}<0$). Density contours have been drawn to help the eye. Only hardness ratios with uncertainties lower than 0.2 are accepted for the sake of reliability.}
\label{L_HRmorph}
\end{figure}

\begin{figure}[t!]
\centering
\includegraphics[width=0.9\columnwidth]{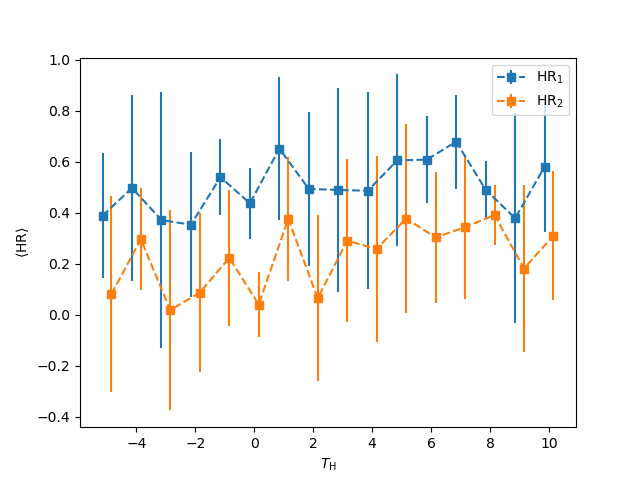}
\caption{$HR_1$ and $HR_2$ across morphological sub-types of host galaxies from \textbf{cULXss} sources. The error bars show the $1\sigma$ dispersion of the values, and the data points have been slightly moved to the left ($HR_1$) and to the right ($HR_2$) to ease visibility.}
\label{HR1and2}
\end{figure}

There are also noteworthy spectral differences within the ULX sample itself. Table~\ref{GalaxyHR} showcases a division between ULX spectral properties in relation to their host galaxies for the \textbf{cULXss}. ULXs in lenticular and elliptical galaxies present the softest spectra in all cases both for faint and bright ULXs. The hardest spectra are found in late spiral galaxies. This notion is clearer in Figure~\ref{L_HRmorph}, where the hardness ratio distributions of ULXs in late-type and early-type galaxies are seen to cluster around slightly different values. However, Figure~\ref{HR1and2} illustrates that the data comes with a large dispersion, despite the increasing trends in the values of $HR_1$ and $HR_2$ with the morphological sub-class of each galaxy up until the irregulars ones. This information points towards photoelectric absorption being higher in spiral galaxies than in elliptical and irregular ones, in agreement with what has been reported in previous \textit{XMM-Newton}-based works \citep{Walton2011,Earnshaw2019}, but also towards unaccounted factors being more determinant for the individual properties of ULXs.

Looking back at Table~\ref{GalaxyHR}, a division between the \textbf{cULXss} and \textbf{cbULXss} is seen. On average, low-luminosity ULXs present harder spectra than the bright ones. It is unlikely that the host galaxies of bright ULX candidates contain less neutral gas in general, and perhaps their generally softer spectra is an indication of bright candidates having a different physical nature than the rest of the candidates. Their emission could also result from the blending of lower-luminosity sources as suggested in Section~\ref{resolution}, perhaps with the diffuse emission of the stellar population in the host galaxies. Within the \textbf{cbULXss} itself, it also appears that candidates in late-type galaxies cluster around slightly harder spectra than those hosted by early-type galaxies, as showcased in Figure~\ref{L_HRmorph}.

\begin{table}[h]
 \cprotect\caption[]{\label{variabilityULX}\textbf{Top:} mean value of $\verb|VAR_PROB|$ from \textit{4XMM-DR9s} according to morphology groups. \textbf{Middle:} occupancy of variability of \textbf{cULXss} and \textbf{cbULXss} variability bins (in \% and as defined in Section~\ref{variable}), also split according to the morphological type of their host galaxy. \textbf{Bottom:} mean max to min flux ratio of sources in each variability bin and morphology group.\\}
 \centering
\begin{tabular}{lcccccc}
 \hline
 \hline
  Sample &\multicolumn{6}{c}{\textbf{cULXss}-\textbf{DR9s}} \\
  Subsample          & All  & El.  & Le.  & ESp. & LSp. & Ir.  \\
  \hline
  Size               & 147  & 28   & 15   & 51     & 49     & 4    \\
  $\langle\verb|VAR_PROB|\rangle$ & 0.18 & 0.23 & 0.26 & 0.20 & 0.10 & 0.22 \\
  \hline
  1$\sigma$ constant & 8.8  & 3.6  & 20.0 & 13.7   & 4.1    & 0.0  \\  
  uncertain          & 15.0 & 28.5 & 13.4 & 9.9    & 8.2    & 50  \\
  1$\sigma$ variable & 15.6 & 21.4 & 13.3 & 19.6   & 12.2   & 0.0  \\
  2$\sigma$ variable & 15.0 & 17.9 & 20.0 & 13.7   & 14.3   & 0.0  \\ 
  3$\sigma$ variable & 45.6 & 28.6 & 33.3 & 43.1   & 61.2   & 50  \\
  \hline
  $\langle\verb|FRATIO|\rangle$   & 72.8 & 6.02 & 4.10 & 26.4 & 186.1 & 2.42 \\
  1$\sigma$ constant & 1.57 & 1.31 & 1.69 & 1.57   & 1.53   & ...   \\
  1$\sigma$ variable & 11.4 & 2.04 & 2.50 & 22.82  & 3.51   & ...   \\
  2$\sigma$ variable & 10.0 & 4.39 & 6.24 & 3.04   & 2.61   & ...   \\
  3$\sigma$ variable & 151.6& 14.5 & 5.87 & 48.8   & 297.7  & 3.70 \\
\hline
 \hline
  Sample &\multicolumn{6}{c}{\textbf{cbULXss}-\textbf{DR9s}} \\
  Subsample          & All  & El.  & Le.  & ESp. & LSp. & Ir.  \\
  \hline
  Size               & 31   & 3    & 8    & 12     & 7      & 0    \\
  $\langle\verb|VAR_PROB|\rangle$ & 0.42 & 0.41 & 0.35 & 0.39 & 0.49 & ... \\
  \hline
  1$\sigma$ constant & 22.6 & 0.0  & 12.5 & 25.0   & 28.6   & ...   \\
  uncertain          & 35.4 & 66.7 & 37.5 & 41.7   & 42.8   & ...   \\
  1$\sigma$ variable & 19.4 & 33.3 & 25.0 & 25.0   & 0.0    & ...   \\
  2$\sigma$ variable & 3.2  & 0.0  & 0.0  & 8.3    & 0.0    & ...   \\
  3$\sigma$ variable & 19.4 & 0.0  & 25.0 & 16.7   & 28.6   & ...   \\
  \hline
  $\langle\verb|FRATIO|\rangle$  & 11.6 & 2.90 & 3.29 & 2.40 & 42.0 & ... \\
  1$\sigma$ constant & 1.95  & ...   & 5.55 & 1.53 & 1.09  & ...   \\
  1$\sigma$ variable & 1.85  & 1.90 & 1.42 & 2.12 & ...    & ...   \\
  2$\sigma$ variable & 4.31  & ...   & 4.31 & ...   & ...    & ...   \\
  3$\sigma$ variable & 49.92 & ...   & 4.57 & 3.67 & 141.5 & ... \\
\hline
\end{tabular}
\end{table}

\subsection{ULX variability}\label{variable}

The differences between ULX candidates also extend to their variability properties. We find 147 \textbf{cULXss} counterparts within three times their positional uncertainty in the \textit{4XMM-DR9s} objects with \verb|N_CONTRIB|$\geq2$. This number goes down to 31 for the \textbf{cbULXss} sample. For all these sources, we have at our disposal variability information drawn from their multiple overlapping observations \citep{Traulsen2019}.

We pay attention to \verb|VAR_PROB|, which is the probability of a source \textit{not} showcasing variability between observations, in essence a long-term variability measure. For our analysis, we put the ULX candidates in five bins of variability probability. We consider objects with $\verb|VAR_PROB|>1\sigma$, where $1\sigma=0.6827$, to be most likely constant, and we collect them in the \textit{$1\sigma$ constant} bin. Thereafter, we collect the objects with $1-1\sigma>\verb|VAR_PROB|>1-2\sigma$, $1-2\sigma>\verb|VAR_PROB|>1-3\sigma$ and $\verb|VAR_PROB|<1-3\sigma$ into the \textit{$1\sigma$ variable}, \textit{$2\sigma$ variable} and \textit{$3\sigma$ variable} bins, being $2\sigma=0.9545$ and $2\sigma=0.9973$, while the remainder are left in the \textit{uncertain} variability bin. We also register the \verb|FRATIO| values of the matched objects, which corresponds to the ratio between the highest and the lowest detected flux from a source, and compute the average value not only for every population time, but also for every \verb|VAR_PROB| bin.

The results are shown in Table~\ref{variabilityULX}, where it is seen that ULX candidates from the \textbf{cULXss} hosted by late spiral galaxies are the ones with the highest chances of showcasing variability between different observations, as they present the lowest \verb|VAR_PROB| average value, with 61.2\% of them belonging to the \textit{$3\sigma$ variable} bin, and only 4.1\% staying in the \textit{$1\sigma$ constant} bin. In contrast, ULX candidates in lenticular galaxies present the highest \verb|VAR_PROB| average value, falling 20\% of them in the \textit{$1\sigma$ constant} bin. However, the ULX population with the lowest fraction of candidates in the \textit{$3\sigma$ variable} bin is that of candidates hosted by elliptical galaxies. ULX candidates hosted by early spiral galaxies present in-between variability properties.

Regarding the measured variation of fluxes between different observations, Table~\ref{variabilityULX} also shows that for ULX candidates hosted by late-spiral galaxies the \verb|FRATIO| values span two orders of magnitude on average, both for the entire set and for those in the \textit{$3\sigma$ variable} bin. ULX candidates in lenticular galaxies showcase the lowest \verb|FRATIO| values, being followed by those in elliptical galaxies. In these cases, the inter-observation luminosity variability does not go beyond one order of magnitude. ULXs in early spiral galaxies repeat their role as a bridge between late spiral-hosted and lenticular-hosted candidates. Not surprisingly, candidates in the \textit{$1\sigma$ constant} present average values of \verb|FRATIO| of the order of 1 regardless of their host galaxy, and while some exceptions are seen, in general \verb|FRATIO| increases with decreasing \verb|VAR_PROB|.

Some trends are similar for the \textbf{cbULXss}. As shown in the lower half of Table~\ref{variabilityULX}, bright candidates hosted by late spiral galaxies are the ones with the highest fraction of ULX candidates in the \textit{$3\sigma$ variable} bin, and they once again have the highest \verb|FRATIO| average values. However, they are typically less variable than their \textbf{cULXss} counterparts, as seen from the higher \verb|VAR_PROB| and lower \verb|FRATIO| average values for the entire set of candidates regardless of the morphological type of the host galaxy. The difference between early spiral-hosted and elliptical and lenticular-hosted ULXs is also blurrier, most likely due to the small size of the available sample.

It should be noted that despite these general trends, the dispersion in the \verb|VAR_PROB| and \verb|FRATIO| values is too large to find a reliable dependence on other quantities. It is tempting to propose a relationship between variability and the star-formation activity in galaxies, as coincidentally late spiral galaxies also present the highest sSFR -excluding irregular galaxies from the analysis-, while both elliptical and lenticular galaxies present the lowest sSFR values. Despite these hints towards a correlation between variability and galaxy morphology, a further investigation of the variability properties in the context of the sSFR of the galaxies does not show and clear correlation.

\cite{Sutton2013} make a thorough study of the variability of several ULXs, stating that an increase of X-ray brightness usually goes in hand with a softening of the source due to accretion winds narrowing their funnel shape and obscuring the inner regions of the disc. We do not find such relationship for many cases either. For instance, NGC 5204 X-1, one of their notable examples, which appears in our \textbf{cULXss} with $\langle L_\textbf{X}\rangle=(6.0\pm0.6)\times10^{39}$\,\lum, presents the complete opposite behaviour within \textit{XMM-Newton} observations, ranging from $HR_1\approx0.33$ and $HR_2\approx-0.02$ at $L_\textbf{X}\approx5\times10^{39}$\,\lum\ to $HR_1\approx0.42$ and $HR_2\approx0.02$ at $L_\textbf{X}\approx8\times10^{39}$\,\lum. Nonetheless, the methodology and sample data in both of our studies diverge greatly, and in particular our methodology is rather limited when making statements about individual sources.

\section{Discussion}\label{discussion}

\subsection{Limitations of our study}

\subsubsection{Limitations of the catalogue}

With 779 identified candidates, our catalogue provides the largest ULX sample built up to date, only followed by the 629 candidates provided by the \textit{Chandra}-based catalogue from \cite{Kovlakas2020}, the 470 candidates from \cite{Walton2011} and the 394 candidates from \cite{Earnshaw2019}. Out of these, 94 are bright candidates with at least one detection with $L_\textrm{X}>5\times10^{40}$\,\lum. However, to avoid biases towards bright sources and blending, we have only considered the 292 candidates from the \textbf{cULXss} and 62 from the \textbf{cbULXss} for the population study. While these ensures the veracity of our results for the whole \textbf{cULXss}, the small size of the \textbf{cbULXss} hampers or study severely. In addition, as suggested in Section~\ref{IMBH}, the bright population is more prone to contamination by unidentified interlopers and source blending, hampering the study even further.

Aside from the manual inspection of sources during the filtering process, our study is also strictly limited to collective properties of our samples. Both the interloper and the ULX populations are very heterogeneous samples at the astrophysical level, and while some trends are identified, individual objects can differ greatly from each other. For instance, we have identified in Section~\ref{spectral} that ULX candidates in late-type galaxies tend to present harder spectra than those hosted by early-type ones. However, it is postulated that ULXs present hard or soft states depending on the viewing angle and the accretion rate \citep{Sutton2013}, and therefore it is most likely that the spectrum of an individual ULX is determined by these factors rather than the morphological type of their host galaxy. 

From a population study such as the one presented in this work we cannot draw any conclusions on the nature of individual sources. However, we can make connections between their collective properties (spectra, variability, frequency) and the nature of the stellar populations in their host galaxy. For example we see a trend for ULXs in late type spirals to exhibit harder colours, indicative of more significant photoelectric absorption. However, we do not see any trend of variability indicators with the sSFR.  The variability and spectral trends can be better addressed with systematic observations of individual objects \citep[e.g.,][]{Sutton2013,Kaaret2017}.

\subsubsection{Selection biases}

\begin{figure}[t!]
\centering
\includegraphics[width=\columnwidth]{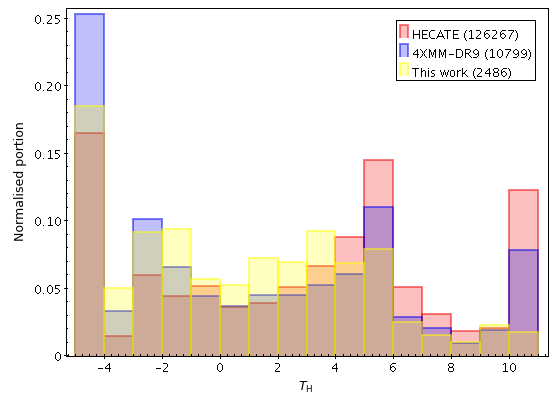}
\caption{Hubble Type distributions of galaxies with listed values in \textit{HECATE}, within \textit{4XMM-DR9} \textit{XMM-Newton} observation, and within our catalogue. Portions have been normalized to sum 1 in all of the three cases due to the large differences in size between the sets.}
\label{HubbleBias}
\end{figure}

The main limitation of \textit{4XMM-DR9} is that it has been constructed from observations of galaxies or clusters that were already of interest to astronomers. As such, the characteristics of our ULX sample may deviate from those of the true ULX population. In Figure~\ref{HubbleBias} it is seen how galaxies included in \textit{4XMM-DR9} present a slight bias towards earlier types with respect to the full content of \textit{HECATE}. This bias only gets amplified further down in the construction of our catalogue.

Choosing to use \textit{4XMM-DR9} as the only X-ray catalogue of reference also limits the size of the X-ray sample available to us, as it covers only 2.7\% of the sky area and 8.6\% of all the galaxies present in \textit{HECATE}. Fortunately, other recent works such as \cite{Kovlakas2020} already take into account non-overlapping parts of the sky, despite suffering the same bias as our own individually.

The most straightforward way to tackle these issues is studying the ULX content of blind all-sky surveys. A good opportunity for this is presented by the still on-going \textit{eROSITA} All Sky Survey \citep[\textit{eRASS},][]{eROSITA2012,predehl+21}. By observing the entirety of the sky instead of selected patches, and accounting for the average \textit{eROSITA}'s sensitivity at the end of the \textit{eRASS}, we assess that around $\sim$1\,000 new ULX candidates will be discovered in \textit{HECATE} galaxies of at least 20\arcsec\ in size, \textit{eROSITA}'s FWHM angular resolution.

Further biases in our catalogue stem from our own design choices, particularly from focusing our studies on source parameters. Occasionally, quality flags and some measures of source parameters pick the worst possible value among all the values assigned to individual detections. This means that if a legitimate source has one detection with $\verb|SUM_FLAG|>1$, this bad value percolates into \verb|SC_SUM_FLAG| and renders the source non-elegible for being a ULX candidate under our criteria. Ignoring the \verb|SC_SUM_FLAG| value, the total number of ULX candidates of quality raises to 847. This choice was made because a bad detection often leads to unreliable source parameters, but the true extent of this effect would need to be checked individually for every source. Nonetheless, all of these sources are still available to the user of the catalogue.

Another bias we introduced was the selection of point-like sources with high detection likelihood, which excluded 1\,632 sources within \textit{HECATE}'s isophotal ellipses, in contrast to the 23\,262 accepted ones in the final catalogue. However, we do think that the exclusion of extended sources is justified, as ULXs are point-like sources.

\subsubsection{\textit{XMM-Newton}'s angular resolution}\label{resolution}

\textit{XMM-Newton}'s angular resolution is a limiting factor due to the clustered nature of star-forming regions, where ULXs tend to be found \citep{Anastasopoulou2016,Kaaret2017}. H II regions, both Galactic and extragalactic, present sizes of the order of $\sim$10~pc, at most $\sim$100~pc \citep[plus references within]{Tremblin2014,HuntHirashita2009}. This implies that we should expect contamination from neighbouring X-ray sources in the ULX candidate parameters at $D\gtrsim10$~Mpc, or even blending of sources into spurious ULX candidates. The \textbf{cULXss} deals with this issue by establishing the cut at $D_\textrm{max}\approx29$~Mpc, so we expect little contamination to our ULX population study. However, it is very telling how at $D_\textrm{max}\gtrsim100$~Mpc the bright ULX candidates completely overtake over the normal ULX candidates. In fact, it was explicitly revealed in Section~\ref{manual} that some of the manually investigated objects suffer from this problem.

Other catalogues built with higher resolution telescopes such as \textit{Chandra} are able to partially alleviate this problem, but the issue reappears at $D\gtrsim40$~Mpc as stated in \cite{Kovlakas2020}. Individual follow up of interesting sources is once again essential to untangle the nature of these objects.

\subsection{Discussion of contents}

\subsubsection{Candidates at the ULX luminosity threshold}\label{ULXthreshold}

As expressed in Section~\ref{ULXcandidates}, our criterion for the selection of ULX candidates consists on whether a source has at least one detection whose luminosity is above this established threshold within its 1$\sigma$ uncertainty. This rather lax condition allows for typically low-luminosity sources to be considered ULX candidates. However, it manages to take into account that some variable objects may behave as ULXs with intermittence, and also NS-ULXs close to the Eddington limit. It is noteworthy that, while in this work we have used the common definition for the ULX luminosity threshold in the literature \citep[$L_\textrm{X}>10^{39}$~\lum,][]{Kaaret2017}, this is but an approximation, as the Eddington limit lies at $1.8\times10^{38}$~\lum~for a 1.4 M$_\odot$ NS, and at $1.3\times10^{39}$~\lum~for a 10 M$_\odot$ StBH. 

As exposed in Section~\ref{contents}, only 516 of our ULX candidates have at least one detection with luminosity above the ULX threshold with more than 1$\sigma$ significance ($L_\textrm{X}-\Delta L_\textrm{X}>10^{39}$\,\lum) and 150 other are also likely to be ULXs within their 1$\sigma$ uncertainty ($L_\textrm{X}>10^{39}$\,\lum). This leaves the catalogue with 106 objects that qualify as ULXs only within the uncertainty of their brightest detection ($L_\textrm{X}<10^{39}$\,\lum), and are therefore unlikely to be ULXs according to the common definition. However, from this sub-set, a vast majority of 95 have been observed with luminosity well above that of the Eddington limit for a 1.4 M$_\odot$ accreting NS ($L_\textrm{X}-\Delta L_\textrm{X}>1.8\times10^{39}$\,\lum), and therefore still deserve a place in this catalogue in consideration of a more comprehensive ULX luminosity threshold.

\subsubsection{The most luminous ULXs}\label{IMBH}

\begin{table}[t]
 \caption[]{\label{unknownContTable}The number of ULX candidates according to their host galaxies, the number of background (AGNs, QSOs and background galaxies) and foreground (stellar objects) that would otherwise qualify as ULX candidates, and the total fraction that they would constitute if added to the total.}
 \centering
\begin{tabular}{lcccccc}
 \hline
 \hline
  Sample &\multicolumn{6}{c}{ULX candidates of quality}    \\
  Subsample     & All & El. & Le. & ESp. & LSp. & Ir.  \\
  \hline
  Candidates    & 779 & 141 & 129 & 261    & 195    & 28   \\
  Background    & 74  & 15  & 13  & 25     & 12     & 3    \\
  Foreground    & 27  & 6   & 3   & 13     & 1      & 1    \\
  Total (\%)    & 11.5 & 13.0 & 11.0 & 22.8    & 6.25    & 12.5  \\
 \hline
 \hline
  Sample &\multicolumn{6}{c}{Bright ULX candidates of quality} \\
  Subsample & All & El. & Le. & ESp. & LSp. & Ir. \\
  \hline
  Candidates & 94 & 15 & 20 & 33 & 14 & 2 \\
  Background & 29 & 7 & 7 & 9 & 3 & 1 \\
  Foreground & 5 & 2 & 0 & 2 & 0 & 0 \\
  Total (\%) & 26.6 & 37.8 & 26.9 & 25.0 & 17.6 & 33.3 \\
  \hline
\end{tabular}
\end{table}

Our catalogue contains 94 ULX candidates of quality with at least one detection with luminosity $L_\textrm{X}>5\times10^{40}$\,\lum. This sample extends to 109 sources in total if we ignore the \verb|SC_SUM_FLAG| column. It has been exposed in sections~\ref{spectral}~and~\ref{variable} that, while the \textbf{cbULXss} follows the same trends as the \textbf{cULXss}, in the sense that candidates hosted by late-type galaxies tend to present harder spectra and higher variability, it is also true that the bright candidates have generally softer spectra and more subdued variability than their \textbf{cULXss} cousins.

A tempting way to explain this difference is to invoke different physical origins of the sources. Hyperluminous X-ray sources (HLXs) are ULXs with $L_\textbf{X}>10^{41}$\,\lum, and albeit the community has recently shifted towards explaining them as typical X-ray binaries at the high end of the luminosity distribution due to very large mass transfer rates \citep{Bachetti2016,Wiktorowicz2017}, they are still the best objects where to look for BHs of large masses or even IMBH \citep{Kaaret2017}. This search is further motivated by first direct confirmation ever of an IMBH in the gravitational wave event GW190521 \citep{Abbot2020a,Abbot2020b}, which is proof that black holes of these masses do exist and are to be considered as potential explanation for the brightest ULXs. Indeed, a handful of ULXs such as ESO 243-49 HLX-1 \citep{Kong2007,Servillat2011,Pasham2014} or NGC 2276-3c \citep{Mezcua2015} constitute promising candidates of being such kind of objects, the three of which are included in our catalogue.

In our catalogue, \object{ESO 243-49 HLX-1} corresponds to source $\verb|SRCID|=202045402010003$, and has an average luminosity of $\langle L_\textrm{X}\rangle=(1.7\pm0.4)\times10^{41}$\,\lum\ and a maximum luminosity of $L_\textrm{X}=(7\pm2)\times10^{41}$\,\lum. Unfortunately, one of its 6 detections has $\verb|SUM_FLAG|=3$, so it has not been included among our list of ULX candidates of quality. \object{M82 X-1} also has been recovered as source $\verb|SRCID|=201122902010001$, but this source is actually a blend of M82 X-1 and the NS-ULX \object{M82 X-2} \citep{Bachetti2014}, as they are only 0.52" apart and are therefore undistinguishable by \textit{XMM-Newton}. Furthermore, this source also has one detection with $\verb|SUM_FLAG|=3$. \object{NGC 2276-3c} ($\verb|SRCID|=200223402010005$) is the only one of these three that appears as bright ULX of quality. However, it is also constituted by three blended sources in \textit{XMM-Newton}, one of them being a genuine IMBH candidate \citep{Mezcua2015}.

Nonetheless, it is known that not all bright ULX candidates need to explained with the presence of accreting massive BHs. For example, another object classified as a bright ULX candidate in our catalogue is NGC 5907 ULX-1, a NS-ULX that has been observed with a luminosity of $L_\textbf{X}>10^{41}$\,\lum\ \citep{Israel2017}.

This implies that there must be other explanations to their different properties. Many times, the reason may not even be astrophysical. As seen in Figure~\ref{distances}, almost all of them are located at $D\gtrsim20$~Mpc, which makes them prone to be the result of blending of several low-luminosity sources. The fact that they are always found alone in their host galaxies is also a strong indication of source blending. This is reinforced by the explicit detection of confused sources among manually inspected objects in Section~\ref{manual}. It is also possible that there is significant contribution of unidentified nuclear sources to our \textbf{cbULXss}. Table~\ref{unknownCont} showcases well that indeed a much larger share of background interlopers was identified for the bright ULX candidates than in the general sample, implying that the unidentified contaminants may also constitute a larger fraction.

Nonetheless, from the forementioned objects, only NGC 2276-3c has $\langle L_\textrm{X}\rangle>10^{41}$\,\lum. In fact, only 25 belong to the HLX class within their uncertainty, 33 if \verb|SC_SUM_FLAG| is ignored. From the ones with $\verb|SC_SUM_FLAG|\leq1$, only three match with the brightest objects from \cite{Earnshaw2019}, the study closest to ours in terms of sampling and methodology. NGC 4077, IC 4596 are included in \cite{Earnshaw2019} as potential IMBH candidates. IC 4320 also appears in \cite{Earnshaw2019} and \cite{Walton2011}, and has already been investigated by \cite{Sutton2012}. The remaining 22 do not seem to appear in previous literature as far as we are aware. If extended to other objects qualified as contaminants, we also find a match for the HLXs in IC 4252 and UGC 6697. The former is considered a central source in our catalogue, while the later holds $\verb|n_Galaxies|=2$, $\verb|SC_SUM_FLAG|=2$ and has been identified with a \textit{PanSTARRS1} counterpart, and therefore they do not meet the requirements for being considered sources of quality. Others also included in \cite{Earnshaw2019}, such as UGC 1934 and NGC 2276 belong to our bright ULX candidates, but do not qualify as HLXs.

\subsubsection{Effectiveness of the filtering pipeline}\label{unknownCont}

In Table~\ref{HRcont}, it is seen that a total number of 1\,943 sources are matched with AGNs, QSOs or background galaxies, and 1\,652 with stellar objects. Additionally, in Table~\ref{unknownContTable} it is shown that 74 and 27 objects in each of this group would have been classified as ULX candidates if it were not for their association. In total, they would constitute a 11.5\% of the ULX candidates of quality. It is therefore natural to ask whether the exclusion of these objects is accurate or, in the opposite end, sufficient at all.

In the case of stellar objects, we have applied extra conditions besides positional coincidences based on the X-ray flux and optical magnitude of the sources. This way, an X-ray source was only classified as a stellar contaminant if the stellar counterpart was bright enough to explain the X-ray emission or, in the least, to contaminate it significantly. And in any case, most of the objects classified as stellar contaminants concentrate in nearby galaxies devoid of ULX candidates. Nonetheless, the best way to attest the efficacy of the pipeline is comparing the spectral values of ULXs to those of stellar contaminants, shown in Tables~\ref{GalaxyHR}~and~\ref{HRcont}, and seeing that the set of \textbf{cULXss} and the \textbf{cbULXss} have distinguishable spectral properties with respect to the bulk of stellar objects.

On the other hand, the direct exclusion of any X-ray source coincident with a known QSO, AGN or \textit{SDSS-DR14} source may rise more concerns. To inspect whether this method has overshot or has been insufficient, we can compare the number of expected background sources to the identified ones. In Section~\ref{filter} it is mentioned that around 670 background contaminants are expected in the set of galaxies at $D>1$~Mpc from the $\log{N}-\log{S}$ presented in \cite{Mateos2008}, which is equal to ${\sim}12$\% of sources at that distance if it were to be true. In our catalogue, 538 sources at $D>1$~Mpc (${\sim}10$\%) were actually classified as possible QSOs or AGNs. Therefore, it can be concluded that our filtering pipeline does a good job on identifying most of the background contaminants, but also that ${\sim}2$\% of our sources may still be unidentified background sources, including ${\sim}15$ ULX candidates.

\subsection{Comparison with other works}

\subsubsection{Previous serendipitous \textit{XMM-Newton} ULX catalogues}

\cite{Earnshaw2019} followed much of the methodology presented in \cite{Walton2011}. As \cite{Walton2011} build their sample from \textit{2XMM-DR1}\footnote{\url{http://xmmssc.irap.omp.eu}} and \cite{Earnshaw2019} from \textit{3XMM-DR4}, their work can be seen as an update on the ULX content of the \textit{XMM-Newton} survey. Likewise, our work can also be seen as an update on the ULX content of \textit{XMM-Newton} with respect to \cite{Earnshaw2019}. Therefore, we expect to recover most of the ULX candidates presented their work and \cite{Walton2011}.

A source positional cross-match recovers 1\,008 sources from the total of the 1\,314 listed in \cite{Earnshaw2019}. As \cite{Earnshaw2019} do not preserve sources classified as interlopers, 80\% of the matches correspond to clean sources in our side. The remaining sources consist of 11 central sources, 30 stars, one QSO, 16 \textit{SIMBAD} objects, 21 objects from \textit{GaiaDR2}, 32 extragalactic objects from \textit{PanSTARRS1} and 1 background galaxy manually identified in \textit{NED}.

\begin{figure}[t!]
\centering
\includegraphics[width=0.9\columnwidth]{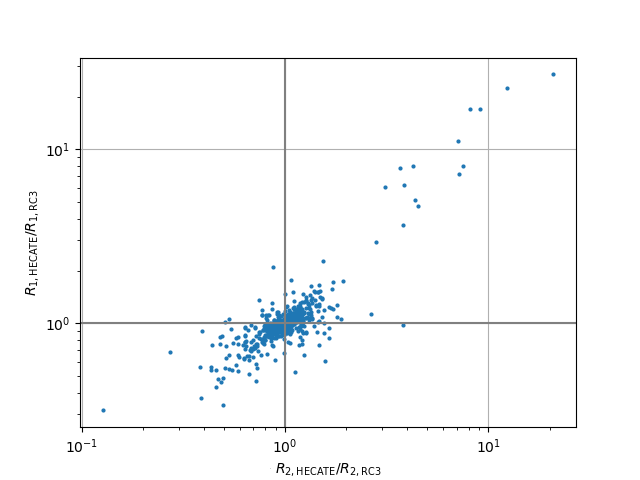}
\caption{Dispersion of the \textit{HECATE} to RC3 semi-major and semi-minor axes ratios for galaxies appearing in both this work and \cite{Earnshaw2019}, respectively.}
\label{HECATE_RC3}
\end{figure}

\begin{figure}[t!]
\centering
\includegraphics[width=0.9\columnwidth]{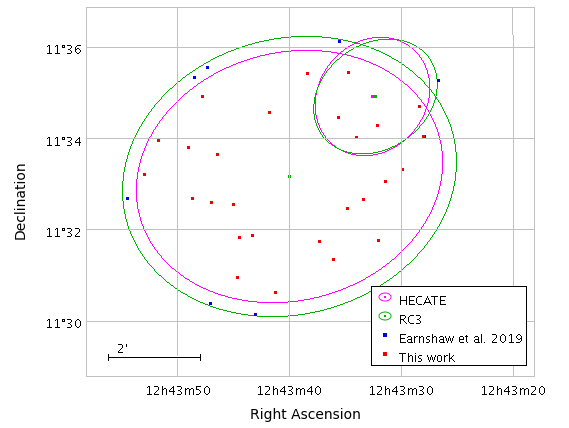}
\caption{TOPCAT astrometric maps of sources found in galaxies NGC 4649 (large ellipses at the center) and NGC 4647 (smaller ellipses at the upper right). The isophotal ellipses from RC3 and \textit{HECATE} are also shown.}
\label{NGC4649and4647}
\end{figure}

The missing 306 sources are explained mostly by the \textit{HECATE} updates to the isophotal radii of galaxies with respect to \textit{RC3} \citep{deVaucouleurs1991,Corwin1994} used in \cite{Earnshaw2019}. Figure~\ref{HECATE_RC3} showcases how most of the newly updated galaxy dimensions often differ slightly from the older values, both towards larger and smaller values. This leads both to the loss and new inclusion of some peripheral sources if the new sizes are smaller or larger, respectively. This phenomenon is well illustrated in Figure~\ref{NGC4649and4647}, where it is seen how smaller isophotal ellipses lead to the exclusion of seven sources from \cite{Earnshaw2019}. For the same reason, we only recover 271 galaxies as sometimes all the sources present in one galaxy are lost. As a rough estimate for the loss of sources, we take 154 galaxies present in both \cite{Earnshaw2019} and our catalogue and that are smaller in \textit{HECATE} than RC3 and compute that, on average, a 26\% of the area sky area is lost. On the other hand, for 112 galaxies that are larger, hold $\verb|n_Galaxies|=1$, and excluding a source that we found to be matched in different galaxies for both catalogues, the mean sky area increment is of a 39\%. This gives an intuition into the 23\% of missing sources from \cite{Earnshaw2019}, but it also indicates that some extra sources have been included. Nonetheless, the newer values considered more accurate due to the increase in photometric data \citep{Makarov2014}.

In addition to this, the discrepancies can be further explained by the improvement of detection algorithms in the \textit{4XMM} editions of the serendipitous \textit{XMM-Newton} catalogues with respect to the \textit{3XMM} editions. As such, a fraction of the spurious detections present in \textit{3XMM-DR4} are expected to be properly flagged or not included in \textit{4XMM-DR9}.

Regarding the ULX content, 260 sources classified by us as ULX candidates in our catalogue have a counterpart in \cite{Earnshaw2019}. This implies the recovery of 68\% of their 384 candidates. This can be explained by the more strict filtering pipeline used in our analysis, in addition to the reasons exposed above. We also recover 337 of the 470 objects in \cite{Walton2011}, 72\% of the total.

Our findings regarding the spectral and abundance properties of the \textbf{cULXss} are a confirmation of the findings from \cite{Walton2011} and \cite{Earnshaw2019}. Both of them find an overabundance of ULXs in spiral galaxies and a clear tendency for late-type galaxy hosted candidates to present slightly harder spectra. In addition, \cite{Earnshaw2019} also find that the spectral properties of the ULX population resemble mostly that of the X-ray population in the range of $10^{38}<L_\textrm{X}<10^{39}$~\lum, while the AGN population presents a slightly different distribution. Finally, we recover five of the HLX candidates in their catalogues.

\subsubsection{Previous serendipitous \textit{Chandra} ULX catalogues}

As of the writing of this paper, \cite{Kovlakas2020} is the most recent and largest \textit{Chandra}-based ULX catalogue. Their work goes in parallel to our own, presenting 629 ULX candidates in 309 galaxies out of 23\,043 sources in 2\,218 galaxies. Therefore, our catalogues complement each other to a great extent. Most remarkably, we use the same reference list of galaxies, \textit{HECATE}, which leads to a very interesting comparison of results.

If we restrict our comparison to mutually shared galaxies at $D>1$~Mpc, their numbers decrease to 11\,359 sources in 849 galaxies. \cite{Kovlakas2020} focused on galaxies with $D<40$~Mpc, and selected as ULX candidates all non-nuclear objects in AGN galaxies, or all objects in non-AGN galaxies, which have $L_\textrm{X}>10^{39}$\,\lum\ and present negative pileup and unreliability flags in \textit{Chandra}. Using those conditions, 341 of these sources consist of ULX candidates in their catalogue. These 849 galaxies contain 3\,130 sources in our catalogue, out of which 457 are ULX candidates on our side. Only 2\,091 sources have a direct counterpart on both sides, including 301 ULX candidates on our side and 144 from \cite{Kovlakas2020}. Finally, only 74 of the ULX candidates coincide on both sides.

Two remarkable discrepancies are noticed. Firstly, the larger density of sources in \cite{Kovlakas2020} in comparison to our own catalogue, and secondly, the lower yield of ULXs in \cite{Kovlakas2020} for the same sky area. The first one can be easily explained by the means of \textit{Chandra}'s sharper resolution, which allows for the detection of many fainter sources which are harder to resolve with \textit{XMM-Newton}. The second point can be explained by the difference of our ULX filtering criterion. We choose as a ULX candidate those objects that, aside from having good data and being clean of interlopers, find themselves within the ULX luminosity regime in at least one of their detections, including the corresponding uncertainties. This implies the inclusion of objects at the ULX luminosity threshold into our ULX candidate set, as discussed in Section~\ref{ULXthreshold}. By contrast, \cite{Kovlakas2020} use only the average source luminosity as their criterion. Using their criterion, our set of ULX candidates in the same area would get reduced from 457 to 347, closer to their 341 candidates. Additionally, we are prone to source blending due to \textit{XMM-Newton}'s poorer resolution, leading to slight luminosity overestimates.

\begin{figure}[t!]
\centering
\includegraphics[width=0.9\columnwidth]{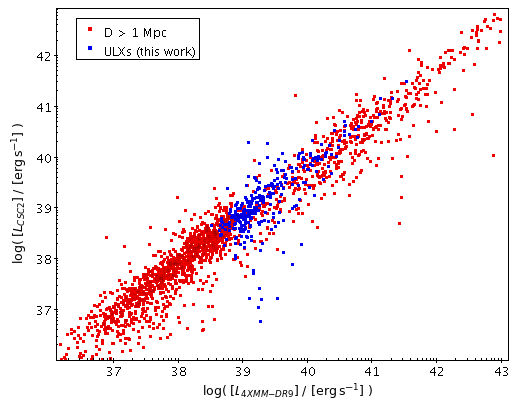}
\caption{Dispersion of source luminosities for coinciding $D>1$~Mpc sources in our catalogue (horizontal axis) and \cite{Kovlakas2020} (vertical axis). ULX candidates in our work catalogue are highlighted in blue.}
\label{USvsKovl}
\end{figure}

There is an extra, more subtle, factor that contributes to the lack of coinciding candidates. As illustrated by Figure~\ref{USvsKovl}, the estimated luminosity of matched sources give are in good agreement between the two catalogues, but with a typical dispersion of around order of magnitude. This alone explains why there are ULX candidates in our side that are not accounted as such in \citep{Kovlakas2020} and vice versa, despite having a positional match on the sky. The source of the spread can be both intrinsic variability in the sources and uncertainties arising during the flux measurement. Nonetheless, these discrepancies add to the mutual completeness of the two catalogues. Even in overlapping observations, both catalogues are to be considered to have a complete view of the total ULX content.

Besides quantity, we perform similar qualitative work using the same parameters as provided by \textit{HECATE}. \cite{Kovlakas2020} find an overabundance of ULX in spiral galaxies as well as a spike in elliptical galaxies, in agreement to our results. The most interesting common result is that the number of ULX candidates per SFR in early-spiral galaxies is the closest to the one predicted in \cite{Wiktorowicz2017}, of 0.67~ULX\psfr. In our case, we estimate a peak value of 0.65~ULX\psfr for early spiral galaxies, while \cite{Kovlakas2020} present a value of 0.51~ULX\psfr. We also agree with \cite{Kovlakas2020} in that there is ULX population in early-type galaxies exists despite their low SFR values, indicating that it is composed by an older population of LMXB. However, regarding the spiral galaxies we see that in the \textit{XMM-Newton} sample galaxies with $T_\textrm{H}\sim3$ present the highest ULX frequencies and fraction of hosts, while \cite{Kovlakas2020} find it at galaxies with $T_\textrm{H}\sim5$ (Sc galaxies in the original paper). This difference will most likely get blurred as ULX catalogue sizes grow in the future.

\subsection{Future prospects}

Our catalogue consists of the largest ULX collection built up to date, and it is based on the most recent \textit{XMM-Newton} serendipitous catalogue available at the time. Nonetheless, the available catalogues keep improving both in size and quality. For instance, in late 2020, the tenth data release of the \textit{XMM-Newton} serendipitous catalogue, \textit{4XMM-DR10}\footnote{\url{http://xmmssc.irap.omp.eu/Catalogue/4XMM-DR10/4XMM_DR10.html}}, was made available to the public, adding 25\,034 new sources from 443 new observations, an increase of almost an extra 5\% with respect to \textit{4XMM-DR9}. As the available samples keep increasing, so will do the samples of ULX candidates available to the community. This will allow for more solid statements regarding the spectral and variability properties of the bright ULX population, and to better identify it as a subgroup of the ULX population.

A particularly promising resource that will soon be available is that of the \textit{eRASS} survey \citep{eROSITA2012}, which has been active for more than a year as of the writing of this paper \citep{predehl+21}. Based on the area observed by \textit{eROSITA} in the western side of the galactic hemisphere as of April 27, 2020, the preliminary study of \cite{Bernadich2020} finds a total number of 132 ULX candidates, of which only 30 are matched to our catalogue within a 3 times their positional uncertainty. This is to be expected, as sources from the \textit{eRASS} in general are spread all across the sky, having the observations little overlap with the \textit{XMM-Newton} field. Taking into account the area covered by \textit{eROSITA} and its average sensitivity at the end of the survey by 2024, the discovery of around 1\,000 new ULX candidates is expected.

X-ray catalogues are not the only ones of relevance to the search for ULXs. Information from other wavelength ranges help to uncover investigate the nature of the observed objects.
A lot of work is being put on that side too. For instance, also in late 2020, the \textit{Gaia Early Data Release 3} was made available to early access, providing positional and apparent brightness information for $10^{8}$ new sources \citep{Gaia2020}, with a full release planned by 2022\footnote{\url{https://www.cosmos.esa.int/web/gaia/release}}. Given our methodology, catalogues like this will be of great aid in identifying further interlopers in future works.

\subsection{Conclusions}

We have built an expanded ULX catalogue from the latest available \textit{XMM-Newton} serendipitous source catalogue, \textit{4XMM-DR9}, and the \textit{HECATE} list of galaxies. A total number of 23\,262 point-like sources of high-detection likelihood have been included within the isophotal ellipses of \textit{HECATE} galaxies, out of which 3\,274 have at least one detection within the ULX luminosity regime ($L_\textrm{X}>10^{39}$~\lum). However, most of these sources consist of contaminating interlopers, 2\,208 of them being the nuclear source of their host galaxy. We have built a filtering pipeline to identify such possible interlopers from other available catalogues and databases, such as \textit{PanSTARRS1}, \textit{Tycho2}, \textit{SDSS-DR14}, \textit{VéronQSO}, \textit{SIMBAD}, \textit{PanSTARRS1} and \textit{NED}. In the end, 779 objects in 617 galaxies qualify as ULX candidates, with 94 with at least one detection with luminosity over the $L_\textrm{X}>5\times10^{40}$\,\lum\ threshold. Around 30 of these objects qualify as HLXs, with $\langle L_\textrm{X}\rangle>10^{41}$\,\lum, many of which require individual follow-ups to untangle their physical nature.

We show that ULX candidates from our complete sub-sample are preferably found in late-type galaxies, in agreement to the notion presented in recent ULX census \citep{Kovlakas2020,Earnshaw2019} and reviews \citep[][and references therein]{Kaaret2017}. In particular, early and late spiral galaxies present the largest ULX frequencies, with 49\% of early spiral galaxies and 37\% late spiral galaxies hosting at least one ULX. These galaxies are also the ones with the highest SFR per unit of stellar mass according to the \textit{HECATE} values, confirming the correlation between ULX and star-formation. We compute of 0.49 ULX\psfr\ for early spirals and up to 1.01 ULX\psfr\ for late spirals. These numbers range across the values exposed in previous literature \citep{Wiktorowicz2017,Kovlakas2020}.

From the hardness ratio measurements provided by \textit{4XMM-DR9}, we have shown that our ULX sample has spectral properties that are distinct from those of the interloper population. At the same time, we show that ULX candidates hosted by late-type galaxies tend to have harder spectra, most likely due to higher photoelectric absorption. These results are overly similar to those already presented by previous \textit{XMM-Newton}-based ULX studies, such as \cite{Walton2011} and \cite{Earnshaw2019}. The same trend is seen in the brighter candidates, but they have softer spectra in general.

From the stacked version \textit{4XMM-DR9s}, built from overlapping \textit{XMM-Newton} observations, we also see that ULX candidates hosted by late spiral galaxies are the ones with the highest probability of showcasing inter-observation variability, with the highest amplitude modulations, while candidates hosted by elliptical and lenticular galaxies fall on the other end. As far as we are aware, this is the first study of this kind. Other variability studies such as \cite{Sutton2013} focus on a more thorough study of individual objects rather than on the whole sample, and therefore the results are not fully comparable. Indeed, to fully understand the variability properties of a population as heterogeneous as ULXs, individual follow-ups are essential. Again, the bright candidates distinguish themselves by showing milder variability in general, albeit similar trends attempt to surface among them.

We are not able to make solid statements regarding the spectral and variability properties of bright candidates due to the small size of the complete bright sub-sample. The expected growth of available X-ray samples in the future will mitigate this problem. Of particular promise is the all-sky blind \textit{eRASS} survey being performed by the \textit{eROSITA} observatory \citep{predehl+21}.

\begin{acknowledgements}

This work was supported by the German DLR under project 50OX1901. 

Konstantinos Kovlakas and Andreas Zezas acknowledge support from the {\it European Research Council} under the European Union's {\it Seventh Framework Programme} (FP/2007-2013) / {\it ERC} Grant Agreement n.~617001, and the {\it European Union’s Horizon 2020} research and innovation programme under the {\it Marie Sk\l{}odowska-Curie RISE} action, Grant Agreement n.~691164 ({\it ASTROSTAT}).

The authors would also like to thank all peers who helped in the construction of the catalogue by offering their own tools and experience, and the anonymous referee who helped to improve this paper.

This research made use of the cross-match service provided by CDS, Strasbourg.

\end{acknowledgements}

\bibliography{ulx.bib}

\end{document}